\documentclass[aps,prb,citeautoscript,superscriptaddress,reprint,twocolumn]{revtex4}

\usepackage{graphicx}
\usepackage{amsfonts,amsmath,amssymb,amsthm}
\usepackage{epstopdf}
\usepackage{upgreek,xspace}
\usepackage{chngcntr}
\usepackage[colorlinks,citecolor=blue]{hyperref}
\def\Mu{\ensuremath\upmu\xspace}

\newcommand {\MZOx}{Mg$_x$Zn$_{1-x}$O}

\newcommand {\MZOZO}{MgZnO/ZnO}

\newcommand {\density}{$\times$10$^{10}$ cm$^{-2}$}

\newcommand {\Bz}{B_\mathrm{z}}
\newcommand {\Bx}{B_\mathrm{x}}

\newcommand {\rs}{r_{\mathrm{s}}}

\newcommand {\nc}{n_\mathrm{c}}
\newcommand {\Bc}{B_\mathrm{c}}
\newcommand{\be}{\begin{equation}}
\newcommand{\ee}{\end{equation}}

\begin{document}

\title{Competing correlated states around the zero field Wigner crystallization transition of electrons in two-dimensions}

\author{J.~Falson}
\email{falson@caltech.edu}
\affiliation{Max-Planck-Institute for Solid State Research, D-70569
Stuttgart, Germany}
\affiliation{Department of Applied Physics and Materials Science, California Institute of Technology, Pasadena, California 91125, USA.}
\affiliation{Institute for Quantum Information and Matter, California Institute of Technology, Pasadena, California 91125, USA}

\author{I.~Sodemann}
\affiliation{Max-Planck-Institute for the Physics of Complex Systems, 01187 Dresden, Germany}
\affiliation{Department of Physics and Astronomy, University of California, Irvine, California 92697, USA}

\author{B.~Skinner}
\affiliation{Department of Physics, Ohio State University, Columbus, Ohio 43210, USA}

\author{D.~Tabrea}
\affiliation{Max-Planck-Institute for Solid State Research, D-70569
	Stuttgart, Germany}

\author{Y.~Kozuka}
\affiliation{Research Center for Magnetic and
	Spintronic Materials, National Institute for Materials Science, Tsukuba 305-0047,
	Japan}
\affiliation{JST, PRESTO, Kawaguchi, Saitama, 332-0012, Japan}

\author{A.~Tsukazaki}
\affiliation{Institute for Materials Research, Tohoku University,
	Sendai 980-8577, Japan}

\author{M.~Kawasaki}
\affiliation{Department of Applied Physics and Quantum-Phase
	Electronics Center (QPEC), University of Tokyo, Tokyo 113-8656,
	Japan} \affiliation{RIKEN Center for Emergent Matter Science
	(CEMS), Wako 351-0198, Japan}

\author{K.~von~Klitzing}
\affiliation{Max-Planck-Institute for Solid State Research, D-70569
	Stuttgart, Germany}

\author{J.~H.~Smet}
\affiliation{Max-Planck-Institute for Solid State Research, D-70569
	Stuttgart, Germany}

\begin{abstract}
The competition between kinetic energy and Coulomb interactions in electronic systems can lead to complex many-body ground states with competing superconducting, charge density wave, and magnetic orders. Here we study the low temperature phases of a strongly interacting zinc-oxide-based high mobility two dimensional electron system that displays a tunable metal-insulator transition. Through a comprehensive analysis of the dependence of electronic transport on temperature, carrier density, in-plane and perpendicular magnetic fields, and voltage bias, we provide evidence for the existence of competing correlated metallic and insulating states with varying degrees of spin polarization. Our system features an unprecedented level of agreement with the state-of-the-art Quantum Monte Carlo phase diagram of the ideal jellium model, including a Wigner crystallization transition at a value of the interaction parameter $r_s\sim 30$ and the absence of a pure Stoner transition. In-plane field dependence of transport reveals a new low temperature state with partial spin polarization separating the spin unpolarized metal and the Wigner crystal, which we examine against possible theoretical scenarios such as an anti-ferromagnetic crystal, Coulomb induced micro-emulsions, and disorder driven puddle formation.
\end{abstract}

\flushbottom
\maketitle

\begin{figure}[hb]
	\includegraphics[width=85mm]{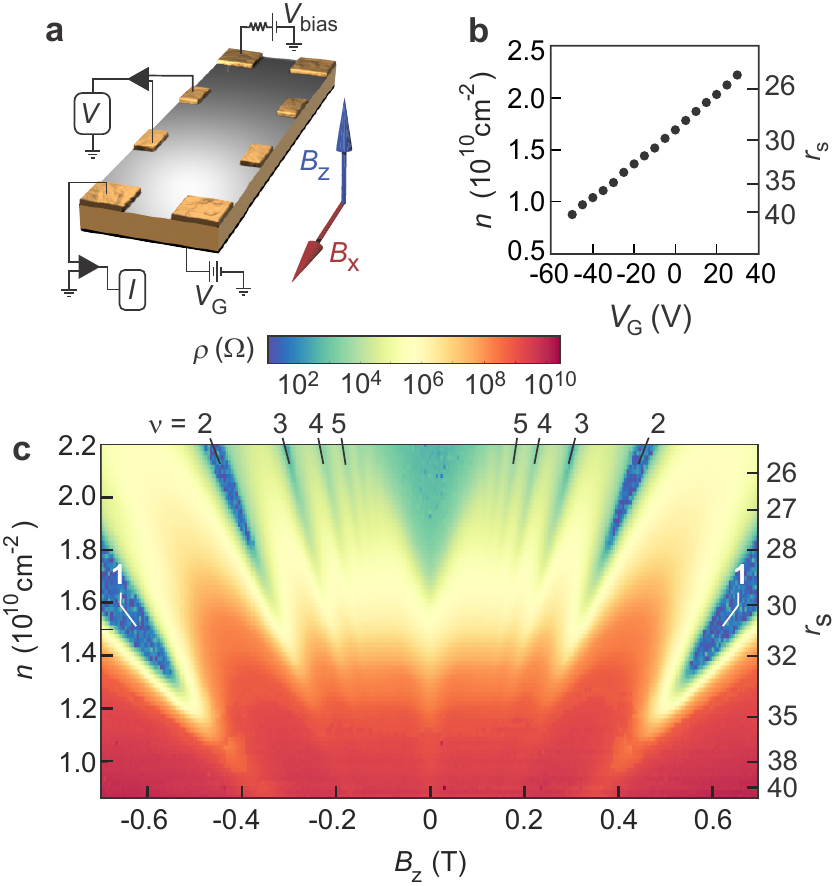}
	\caption{\textbf{The device and quantum transport.} \textbf{a} Schematic of the device under study. \textbf{b} Transfer characteristics of the field-effect operation. \textbf{c} Mapping of transport as a function of $\Bz$~and $n$, taken in the limit $I\rightarrow0$~nA where $T\approx10$~mK. Integer quantum Hall filling factors $\nu$ are noted.}
	\label{Fig1}
\end{figure}

Dilute interacting electrons harbor competing ground states when their Coulomb repulsion greatly exceeds their kinetic energy. In a parabolically dispersing two dimensional electron system (2DES) the ratio of interaction to kinetic energy scales is parameterized by the dimensionless parameter $\rs$, given by
\begin{equation}\label{Eq:rs}
r_{\mathrm{s}} = \frac{1}{(\pi n)^{1/2}a_\mathrm{B}}.
\end{equation}
\noindent
Here, $a_\mathrm{B}= 4\pi \epsilon \hbar^2/m^*e^2$ is the effective Bohr radius of carriers and $n$ is the electron concentration. As the density is lowered, the electron system undergoes a Wigner crystallization transition, which Quantum Monte Carlo (QMC) studies predict to occur at around $\rs \approx 30$ [\onlinecite{tanatar:1989,rapisarda:1996,phillips:1998,chamon:2001,attaccalite:2002,spivak:2004,drummond:2009}]. In spite of decades of research efforts,\cite{spivak:2010,abrahams:2001,Kravchenko:2003,shashkin:2019,dolgopolov:2019} many aspects of the phase diagram of a strongly interacting 2DES in the limit of zero temperature and zero magnetic field remain clouded in the range of {$25< \rs <40$}, where QMC calculations predict a breakdown of the Fermi liquid (FL) state. One of the main obstacles has been the trade-off of interaction and disorder strengths in these platforms; namely, the cleanest systems, such as electron-doped GaAs, are also typically the ones that are relatively weakly interacting, while those with stronger interactions tend to be more disordered. Thus, systematic experimental studies in the high $\rs$ regime ($\rs \geq 20$) remain few.\cite{yoon:1999,knighton:2018,hossain:2020} The advent of ZnO heterostructures, however, offers a new platform that is sufficiently strongly interacting and clean. This combination is evidenced by its display of some of the most fragile correlated states of the fractional quantum Hall regime, such as the 5/2 and 7/2 incompressible states, bubbles and stripes,\cite{falson:2015a,falson:2018b} while still remaining strongly interacting at zero magnetic field, as we demonstrate in this study.

The enhanced electronic interactions in \MZOZO~heterostructures stem primarily from the relatively heavy band mass ($m_\mathrm{b} = 0.3m_0$) and small dielectric constant ($\epsilon=8.5\epsilon_0$). Moreover, the occupation of a single electron pocket at $\Gamma$ combined with weak non-parabolicity and spin-orbit interaction ensure that the system is very close to the ideal jellium model studied in QMC. The bands are highly spin degenerate, and the band $g_\mathrm{b}$-factor ($\approx2$) is isotropic.\cite{kozuka:2013} The quasi-Hall bar device under study is rendered in Fig.\ref{Fig1}\textbf{a}. The epitaxial \MZOZO~heterostructure confines a 2DES approximately 500\,nm beneath the wafer surface with $n$ tuned \textit{in-situ} via a capacitively coupled gate electrode on the back-side of the wafer. The field-effect transfer characteristics are displayed in Fig.\ref{Fig1}\textbf{b}. Here, $n$ is determined from the period of quantum oscillations and Hall effect. Great effort has been invested to perform experiments at very low temperatures, as many transport features are revealed only below $T=50$\,mK. To this end, the sample is immersed within a liquid $^3$He bath in a cryostat that operates down to $T\approx7$\,mK. The electrical characteristics are probed in a four-point configuration by sweeping the DC bias ($V_\mathrm{bias}$) applied while measuring the current across the device ($I$) and local longitudinal voltage drop ($V$), yielding a single $I-V$ trace. The first derivative of this data provides the differential resistance ($dV/dI$) of the device as a function of $I$. We define $\rho$ to be $dV/dI$ in the small current limit ($I\rightarrow0$\,nA), which probes the linear response of the equilibrium state of the system. 

The magnetotransport of the device in the ($\Bz$,$n$) parameter space is presented in Fig.~\ref{Fig1}\textbf{c}. Oscillatory features in $\rho$ as $|\Bz|$ is increased are associated with integer steps in Landau quantization. These states are labeled according to their filling factor $\nu=hn/e\Bz$, where $h$ is the Planck constant and $e$ is the elementary charge. Low-field oscillations begin at approximately 0.07~T, yielding a conservative estimate of the quantum lifetime of carriers $\tau_\mathrm{q}>$10~ps, using the relationship $\tau_\mathrm{q}^{-1}=2\omega_\mathrm{c}$ where $\omega_\mathrm{c}$ is the cyclotron frequency at the onset of oscillations and the band mass. 

\begin{figure}[hb]
	\includegraphics[width=85mm]{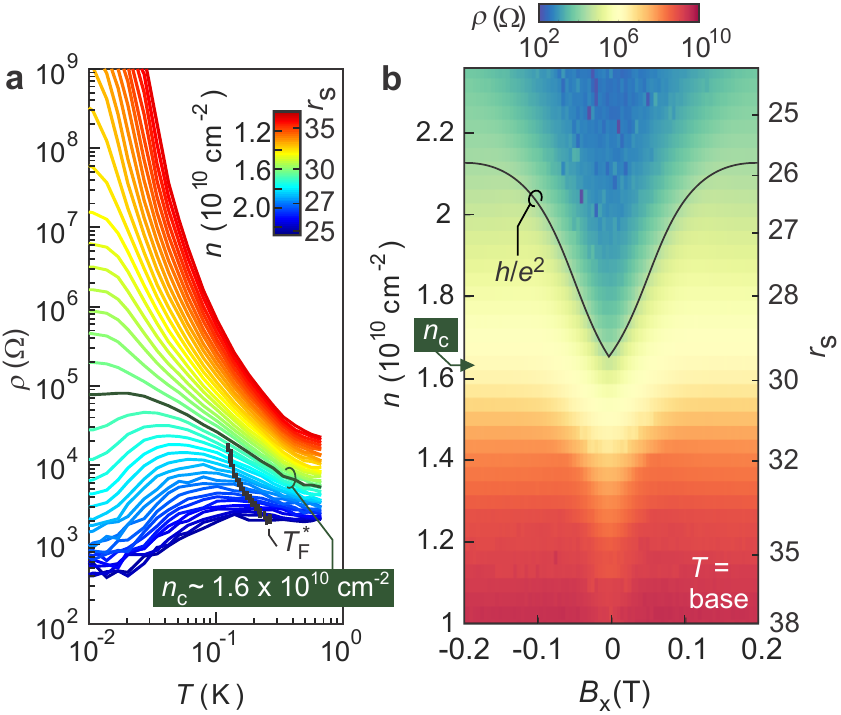}
	\caption{\textbf{The metal insulator transition.} \textbf{a} MIT in the temperature dependence of the resistance as $n$ is tuned. The trace associated with $\nc\approx$1.6\density~is plotted as a bold line. \textbf{b} Transport in the ($\Bx$,$n$)-plane for the device. The condition for $\rho=h/e^2$ is indicated by the black line.}
	\label{Fig2}
\end{figure}

The data in Fig.~\ref{Fig2}\textbf{a} presents $\rho(n,T)$ of the device at zero magnetic field. This data reveals a crossover from a metallic $d\rho/dT>0$ to insulating dependence $d\rho/dT<0$ at a critical density $\nc \approx 1.6$\density~(corresponding to $\rs = 30$), enabling us to associate the density $\nc$ with a zero-field metal-insulator transition (MIT) close to the quantum resistance value $h/e^2$. Data for $T\lesssim 20$~mK deviates from the systematic behavior at higher temperatures, most likely due to the commonly encountered issue of decoupling of the electron temperature from that of the immersion cryogen. The effect of an in-plane magnetic field is displayed in Fig.\ref{Fig2}\textbf{b}, which indicates a positive magnetoresistance at all values of the electron density. The value $\rho=h/e^2$ is identified as a black line, which corresponds to a finite $\Bx$ when $n$ is slightly larger than $\nc$.

\begin{figure}[t]
 	\includegraphics[width=85mm]{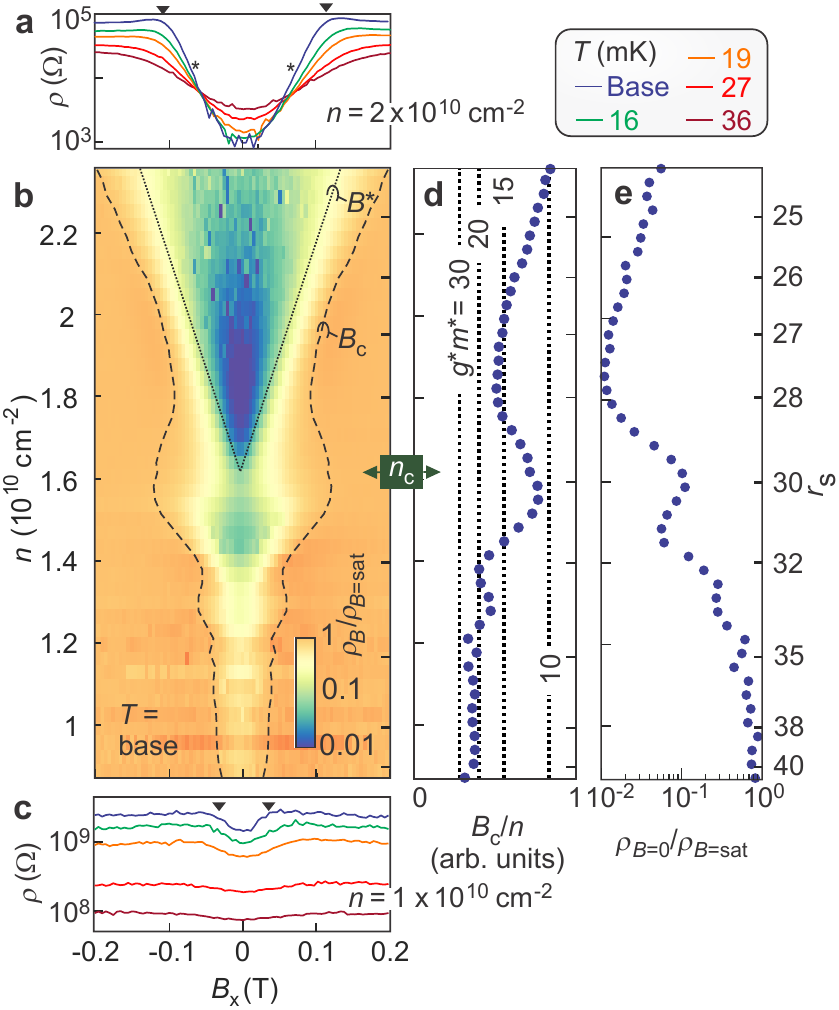}
 	\caption{\textbf{Temperature dependence and examination of spin polarization with application of an in-plane magnetic field.} \textbf{a} $\rho$~as a function of $n$ at distinct temperatures, identifying the critical density $\nc$. \textbf{b} Temperature dependent in-plane magnetotransport when $n$=2\density~with $\Bc$~(black triangle) and $B^*$ ($^*$)  identified. \textbf{c} Mapping of the normalized differential resistance $\rho_B$/$\rho_\mathrm{B=sat}$ as a function of $\Bx$~and $n$. The values of $\Bc$~and $B^*$ are indicated by dashed and dotted lines, respectively. Here, $T\approx10$~mK. \textbf{d} Temperature dependent magnetotransport where $n$=1\density. \textbf{e} $\Bc/n$ as a function of $n$ at base temperature. Corresponding values of $g^*m^*$ are indicated as vertical dotted lines.}
 	\label{Fig3}
\end{figure}

The in-plane magnetic field permits us to directly control the degree of spin polarization of the electrons, as the orbital coupling to in-plane fields is negligible due to the two-dimensional confinement. Figure \ref{Fig3}\textbf{a} plots the magnetoresistance of the device as a function of $\Bx$ at $n=2$\density and at various temperatures. From these curves we can identify two values of the magnetic field of interest: $\Bc$, at which $\rho$ saturates to a value of $\rho_\mathrm{B=sat}$ (black triangle); and $B^*$, at which there is a change in the sign of $d\rho/dT$ from metallic-like to insulating-like. $\Bc$ can be interpreted as the critical field required to reach full spin polarization.\cite{spivak:2010,abrahams:2001,Kravchenko:2003,shashkin:2019,dolgopolov:2019} Figure~\ref{Fig3}\textbf{b} plots the ratio $\rho_\mathrm{B}/\rho_\mathrm{B=sat}$ in the ($\Bx$,$n$)-plane, with orange regions associated with a fully spin polarized 2DES. The data in Fig.~\ref{Fig3} reveals a non-monotonic dependence of $\Bc$~as a function of $n$ (dashed line) as the MIT is crossed. We observe an inflection point in the value of $\Bc/n$ around {$n=1.8$\density}, which is higher than the the value $\nc$ associated with the zero field MIT. The in-plane field traces reveal the presence of finite magnetoresistance even in the low-density limit (Fig.~\ref{Fig3}\textbf{c}), where the device is insulating for all $B$. 

The dotted line in Fig.~\ref{Fig3}\textbf{b} marks the value of $\Bx$~at which the sign of the temperature dependence of resistivity changes, from metallic-like (low $\Bx$) to insulating-like (high $\Bx$). Thus, the ($\Bx$,$n$) parameter space hosts two regions with an insulating-like temperature dependence $d\rho/dT<0$, namely at $B>B^*$ when $n>\nc$, and at all $\Bx$\ when $n<\nc$. As we show in the SI (Sec.\ \ref{sec:SItemperaturedependence}), in the latter regime of $n <\nc$ the temperature dependence is consistent with the activated or VRH mechanisms that are characteristic of insulators \cite{shklovskii_electronic_1984} (including Wigner crystals\cite{shklovskii:2004}). In contrast, in the regime of $n >\nc$ and $B > B^*$ where we encounter what appears as a field-induced MIT, the dependence of resistivity on temperature is more consistent with a linear or power-law relation. Such a linear increase in $\rho$ with $T$, and the accompanying change in sign of $d\rho/dT$, has been shown theoretically to arise in metallic, correlated states with high spin polarization \cite{Zala2001rapid, Zala2001interaction}. Thus, the temperature dependence points to a ground state at $n > \nc$ and $B >\Bc$ that is distinct from the low-density insulating phase for $n<\nc$ at $\rs >30$.

The value of $\Bc$~allows us to measure the renormalized spin susceptibility of the system, $\chi$. Experimentally, it is convenient to use the relationship:
\begin{equation}
B_c \approx \frac{2 \pi \hbar^2 n}{\mu_B g^*m^*}.
\label{Eq.Bc}
\end{equation}
Here, $\mu_B$ is the Bohr magneton, with $g^*m^*$ closely related to the renormalized susceptibility $g^*m^* = 4 \pi g_0 \chi$ (see supplementary section \ref{sec:SIFLparameters} for discussion). The renormalized value of $g^*m^*$ is presented in Fig.~\ref{Fig3}\textbf{d} as a function of $n$. As is observed from much higher densities (see Fig.~\ref{S_gm}), $g^*m^*$ increases monotonically with decreasing $n$ in the range $2.3>n>1.8$\density. The estimated value of $g^*m^*$ at $n=\nc$ represents a nearly 30-fold enhancement over the band value of 0.6. Monitoring the relative magnitude of the magnetoresistance with in-plane field, $\rho_{B=0}/\rho_{B=sat}$, it is evident that the in-plane magnetoresistance is generally suppressed when reducing $n$ (Fig.\ref{Fig3}\textbf{e}). However, the non-trivial dependence of the magnetoresistance never disappears completely, as we discuss below in depth.

\begin{figure*}[ht]
	\includegraphics[width=170mm]{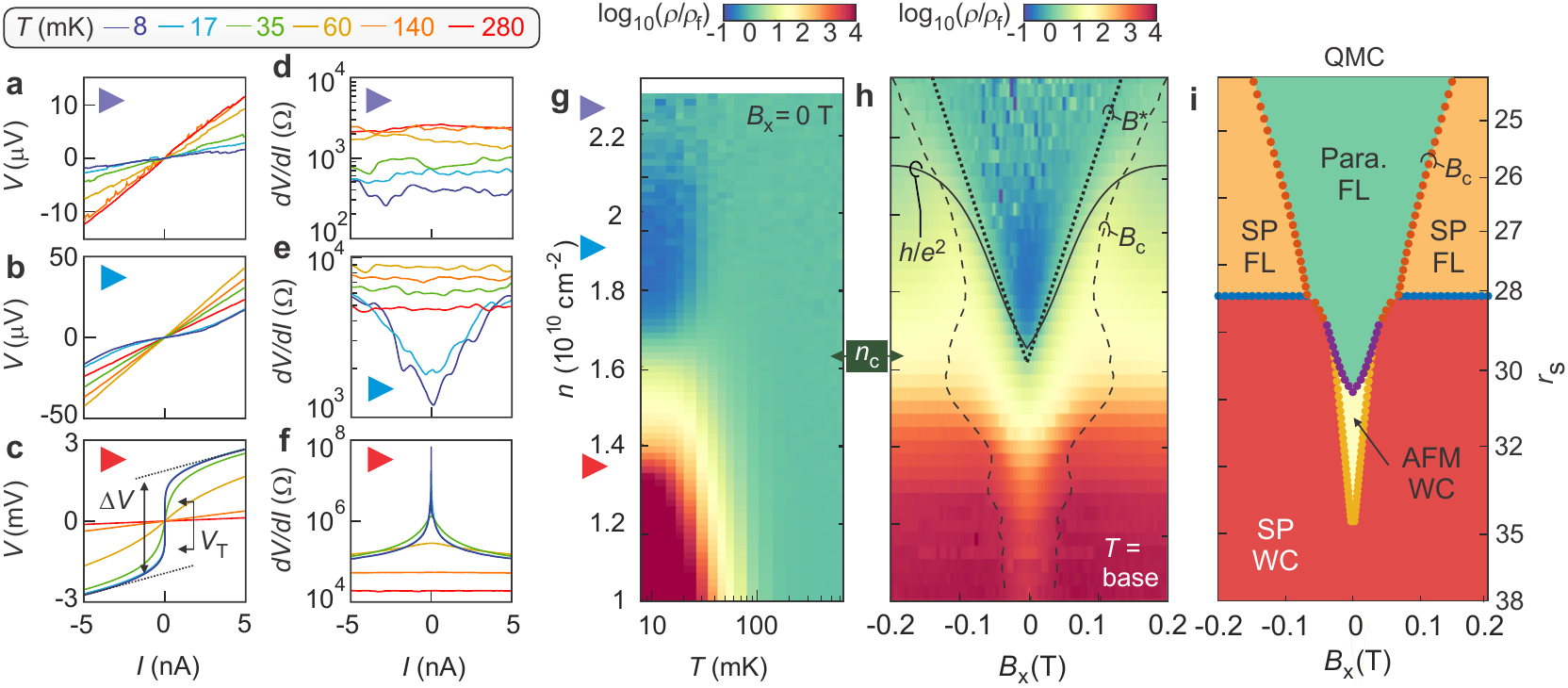}
	\caption{\textbf{Non-linear transport characteristics and phase diagram.} \textbf{a-c} Temperature dependent $I$-$V$ sweeps when $\rs\approx 25$, 28 and 32 and $\Bx=0$\,T. Note the schematic representation of $V_\mathrm{T}$ and $\Delta V$ in panel \textbf{c}. \textbf{d-f} $dV/dI$ of these data sets. Colored triangles highlight the three regimes discussed in the text. \textbf{g}-\textbf{h} Degree of non-linearity, defined as $\rho/\rho_\mathrm{f}$, plotted in \textbf{g} as a function of $n$ and $T$ at $B=0$, and in \textbf{h} as a function of $n$ and $\Bx$~at base temperature. The dashed and dotted represent the previously identified $\Bc$~and $B^*$. \textbf{i} QMC ground states\cite{drummond:2009} with the addition of a finite $B$-field.}
	\label{Fig4}
\end{figure*}

Non-linear charge transport is encountered throughout the parameter space and is revealed by studying the differential resistance as a function of current. Figures \ref{Fig4}\textbf{a-c} plot $I-V$ traces at three distinct charge densities, corresponding to $\rs\approx25$, 28 and 32, respectively (purple, blue and red triangle). The corresponding differential resistance as a function of $I$ is plotted in panels \textbf{d-f}. The three values of $\rs$ represent qualitatively distinct responses in the ($n,T,V_\mathrm{xx}$)-parameter space. For $\rs\approx25$ the system displays metallic ($d\rho/dT>0$) transport as the lowest attainable temperature is approached, with $dV/dI$ approximately constant as a function of $I$. In contrast, strong non-linearity develops in the voltage response as $T\rightarrow 0$ when $n<n_\mathrm{c}$. We characterize this nonlinearity in two different ways: through the threshold voltage ($V_\mathrm{T}$) at which 10~pA flows through the device, and through an extrapolation of the high current voltage drop to zero current ($\Delta V$), as shown in panel \textbf{c}. A large $V_\mathrm{T}$ is a characteristic transport feature expected from a WC ground state,\cite{knighton:2018} arising from pinning of the crystal. We also identify a regime of apparent excess conductance at low bias for a finite range of densities, $\nc <n<2$\density~(panels \textbf{b,e}). In this regime, a flattening of the voltage drop as a function of current is visually apparent in the raw $I-V$ data (Fig.~\ref{Fig4}\textbf{b}), producing a lower differential resistance $dV/dI$ as $I\rightarrow 0$. The differential resistance increases by as much as $4$ times when even a few nanoamperes of current (corresponding to $\sim 1$\,fW power dissipation) are fed through the device. This excess conductance is discussed further below.

Figures \ref{Fig4}\textbf{g} and \textbf{h} plot the ratio of zero-bias and finite-bias resistances, $\rho/\rho_\mathrm{f}$, in the ($n,T$)-plane at zero field (Fig.~\ref{Fig4}\textbf{g}), and in the ($n,B_\mathrm{x}$)-plane at base temperature (Fig.~\ref{Fig4}\textbf{h}). Here, $\rho_\mathrm{f}$ corresponds to the differential resistance at a finite current of approximately 5~nA. The green regions (for which $\log(\rho/\rho_\mathrm{f})$ is close to zero) in these maps correspond to a linear response, for which the differential resistance is independent of $I$, as is evident for all $n$ when $T\geq100$~mK. Some region of excess conductivity, defined as $\rho_f < \rho$, appears as a small dome-like blue region above $\nc$, disappears above 30~mK, and is suppressed with the application of a $B$-field (see Fig.\ref{S_EC}). Similar $I-V$ features have been identified in previous studies as the MIT is approached.\cite{kravchenko:1996,yoon:1999} Both yellow and red regions corresponds to a peak in $\rho$ at $I=0$~nA, with the the former displaying weaker non-linearity in the form of a finite $\Delta V$ and the latter hosting a prominent $V_\mathrm{T}$ as $I\rightarrow0$.

We will now discuss the underlying nature of the phases encountered. We characterize the phases using the boundaries associated with full spin polarization of the system ($\Bc$, dashed line), the change in sign of $d\rho/dT$ ($B^*$, dotted line), the magnitude of the resistivity relative to $h/e^2$ (black line), and the degree of non-linearity ($\rho/\rho_\mathrm{f}$), all of which are plotted in Fig.\ref{Fig4}\textbf{h}. We couple these experimental results with a comparison to state-of-the-art QMC simulations\cite{drummond:2009}, which have identified a competition between paramagnetic FL (Para.~FL), spin polarized FL (SP~FL), antiferromagnetic WC (AFM~WC) with a stripe-like spin order on a triangular lattice and spin-polarized WC (SP~WC) phases in the $\rs$ range studied in our work. Here we expand the phase diagram produced by QMC to take into account a finite in-plane magnetic field (see SI for details and alternative scenarios). The result is presented in Fig.\ref{Fig4}\textbf{i} and contains no free parameters.

We can confidently associate the zero field metal phase at large $n$ with a paramagnetic FL subjected to increasingly strong interactions as $n$ is reduced towards $\nc$. At finite $\Bx$~and $n>\nc$, however, the change in sign of $d\rho/dT$ at $B^*$ that is traditionally associated with a MIT bears closer resemblance to a linear-in-$T$ correction to the conductivity emerging from strong interactions in a FL.\cite{Zala2001rapid, Zala2001interaction} We also note that magnetic field scale at which the paramagnetic FL is predicted by QMC calculations to become the SP FL (Fig.\ref{Fig4}\textbf{i}) agrees very well with the measured value of $B^*$ (or $\Bc$) without any fitting parameters. Therefore we associate the state that appears for $n>\nc$ and $B>\Bc$ as the spin polarized FL. However, while such a linear increase is consistent with our data (see SI Sec.\ \ref{sec:SItemperaturedependence}), we caution that the theory is not \textit{a priori} applicable to the states with relatively large $r_s$ and large resistivity that we are considering. In line with this concern, it is worth emphasizing that our results in this regime exhibit extreme deviations from the usual weak-coupling metallic conductivity, as evidenced, for example, by the enormous positive magnetoresistance and by a low temperature resistivity substantially higher than $h/e^2$.

Turning our attention to $n<\nc$, the insulating phase has the characteristic transport attributes of a pinned WC, as evidenced by the large value of $V_\mathrm{T}$ that develops at low temperature. The non-linearity in this regime is orders of magnitude larger than that of the SP FL phase discussed above, supporting our hypothesis that the two regimes host distinct phases. The positive magnetoresistance at $n < \nc$ that becomes increasing clear at very low temperatures in Fig.~\ref{Fig3}\textbf{c} remains to be fully understood, although it is apparently consistent with calculations\cite{Matveev1995} that consider the effect of Zeeman splitting of localized states on hopping conduction. The presence of finite magnetoresistance appears to preclude the conclusion that the state is fully spin polarized at low temperature in the range of $\rs$ studied. This is in contrast with a recent study of AlAs~\cite{hossain:2020}, which reported an apparent divergence of the spin susceptibility in the insulating phase based upon a flat magnetoresistance for $n < \nc$ at a measurement temperature of $T \approx 0.3$\,K. The lack of clear spontaneous spin-polarization at very low temperatures in our experiment is in agreement with the fact that the exchange energy scale $J$ associated with spin ferromagnetic ordering of the WC at $n<\nc$, as estimated by QMC, is smaller than $|J| <10$\,mK. The spins of the WC are therefore likely disordered by temperature fluctuations at $\Bx=0$. While we note that delicate hysteretic features in transport are indeed resolved for $n \leq n_c$ which upon first glance could indicate some FM ordering (see Fig.\ref{S_hysteresis1}), we, however, ascribe these features to experimental artifacts associated with heating close to zero field and trapped flux in the superconducting coil, as the estimated coercive field in the presence of magnetostatic fields is $\sim 10^{-7}$~T (see section \ref{sec:S_hysteresisdiscussion}) and hence undetectable in experiment.

Finally, we would like to discuss one of the most remarkable findings of our study, namely the non-monotonicity of the in-plane saturation field $\Bc$ near the zero field MIT. The non-monotonicity of $\Bc$ is a low temperature property of the system and is absent above approximately 30~mK (see Fig.\ref{S_TdepBx}). In agreement with state-of-the-art QMC calculations,\cite{drummond:2009} we find no clear evidence for a Stoner instability of the itinerant liquid; finite magnetoresistance is always present in the metal phase. In contrast, QMC has identified a possible AFM crystal in between the paramagnetic FL and the fully SP WC. By adapting the QMC results from Ref.~\onlinecite{drummond:2009} to include in-plane field (see SI section \ref{sec:SIphasediagram} for details), one obtains the phase diagram shown in Fig.\ref{Fig4}\textbf{i}. However, as we see from this phase diagram, the intermediate AFM does not offer any clear explanation for the non-monotonicity of the critical field $\Bc$ to spin polarize the system. Moreover, as mentioned before, our lowest temperature scale $T\approx20$\,mK is larger than the exchange energy scale $J$, or more precisely it is larger than the energy difference per electron of the FM WC and the AFM WC obtained from QMC, as detailed in SI Sec.~\ref{sec:S_Xchangediscussion}, and thus the spin order of the WC is likely destroyed by temperature fluctuations at $\Bx=0$. We therefore believe that the AFM WC phase found in QMC is not likely to be the origin of the non-monotonicity of $\Bc$ that we observe. 
We note that the QMC employed in Ref.\onlinecite{drummond:2009} is variational in nature and therefore it is always possible that other phases not considered could be behind the non-monotonicity of $\Bc$, such as exotic spin liquid states\cite{chakravarty:1999,bernu:2001} or spin-density-wave ordered states like those suggested by Hartree-Fock studies~\cite{bernu:2011,bernu:2017}.

The role of spatial variation in electron density may also be prominent in the vicinity of $n = \nc$. Even in the cleanest samples it is not possible to completely eliminate the role of disorder, which tends to produce variation in the local electron density. When the average density is very close to $\nc$, such variation causes the 2DES to break up into itinerant and localized regions. (See SI Sec.~\ref{sec:SIdisorder} for a more detailed discussion of disorder-induced density modulation.) And even in the absence of disorder, variation in the local density can arise from Coulomb-frustrated phase separation.\cite{spivak:2004,jamel:2005,spivak:2006,li:2019} While such phase separation may be important near $n = \nc$, it does not provide an obvious explanation for the nonmonotonicity of $\Bc$ as a function of $n$. Still, the phase separation picture can serve as a premise for interpreting the apparent excess conductance presented in Figs.~\ref{Fig4}\textbf{b} and \textbf{e}. When conducting and insulating phases are mixed in nearly equal proportion, then electric current flows predominantly through narrow metallic pathways, which are unusually sensitive to joule heating. The ratio of metallic to insulating regions falls as the MIT is approached, with the MIT signifying the transition to a regime where insulating regions percolate and metallic regions are relegated to disconnected puddles. An extended discussion of this scenario is presented in SI section \ref{heating}. 

In summary, our study provides an unprecedented level of experimental clarity about the phase diagram around the Wigner crystallization transition at $r_s \approx 30$ and very low temperatures. Our data reveals a paramagnetic FL which exhibits a strong renormalization of its spin susceptibility as the critical density $\nc$ is approached, with the spin susceptibility becoming nearly 30 times larger than the band value. At $n < \nc$ the transport becomes strongly nonlinear and exhibits an exponential temperature dependence, both of which features are consistent with a WC. The qualitative and quantitative agreement of our phase diagram with state-of-the-art QMC\cite{drummond:2009} is striking with zero adjustable parameters. The most prominent mystery suggested by our measurements is the possible existence and nature of a WC state with incomplete spin polarization. The non-monotonicity of the magnetic field $\Bc$ required to achieve spin polarization has no obvious interpretation in terms of the QMC phase diagram, and may be associated with either spatial coexistence of different phases or with an as-yet undetermined intermediate phase. The large increase of resistance with in-plane field, which for densities near $\nc$ becomes as large as two orders of magnitude, is also incompletely understood.  QMC calculations suggest that the state at $B > B_c$ and $n$ slightly larger than $n_c$ is a spin polarized FL, but the large value of resistance poses a challenge for understanding this state within traditional paradigms of metallic transport.

\section*{\textit{Methods}}
The heterostructure was grown using ozone-assisted molecular beam epitaxy and consists of a lightly alloyed \MZOx~layer ($x \approx$ 0.001) of 500 nm thickness grown on a homoepitaxial ZnO layer upon single crystal (0001) Zn-polar ZnO substrates.\cite{falson:2016,falson:2018a} The heterostructure has an electron mobility approaching 10$^6$~cm$^2$/Vs in the metallic regime. Ohmic contacts were formed by evaporating Ti (10nm) followed by Au (50nm) on the sample surface. Indium was additionally soldered upon these pads to improve the contact quality. The distance between voltage probes is approximately 1~mm. The sample was immersed in a liquid $^3$He containing polycarbonate cell attached to the end of a cold finger of a dilution refrigerator cryostat (650$\Mu$W cooling power at 120~mK) equipped with a 3-axis (9-3-1~T) vector magnet. The $^3$He cell is based upon the design used in other ultra-low temperature experiments on high mobility 2DES.\cite{pan:1999} The mixing chamber temperature is measured using a calibrated cerous magnesium nitrate paramagnetic thermometer for $T\leq100$~mK, and a ruthenium oxide thermometer for $T\geq100$~mK. Each measurement wire is passed through a large surface area ($A\approx$1~m$^2$) sintered silver heat exchanger to overcome the Kapitza interfacial resistance that suppresses heat exchange at low $T$. The differential resistance data is obtained by measuring $V$-$I$ data using DL Instruments 1211 current and 1201 voltage preamplifiers at discrete steps in the ($B,T,n$) parameter space, followed by differentiation using data analysis software. 

\section*{Acknowledgments}
We appreciate discussions with  Joe Checkelsky, Neil Drummond, Jim Eisenstein, Andrea Young, Steve Kivelson, Boris Spivak, and Chaitanya Murthy, along with technical support from Jian-Sheng Xia, Neil Sullivan and Gunther Euchner. M.K. acknowledge the financial support of JST CREST Grant Number JPMJCR16F1, Japan. Y.K. acknowledges JST, PRESTO Grant Number JPMJPR1763, Japan. J.F. acknowledges support from the Max Planck Institute--University of British Columbia--University of Tokyo Center for Quantum Materials, the Deutsche Forschungsgemeinschaft (DFG) (FA 1392/2-1) and the Institute for Quantum Information and Matter, an NSF Physics Frontiers Center (NSF Grant PHY-1733907).

\section*{}
\bibliographystyle{apsrev}

\clearpage

\onecolumngrid
\newpage
\begin{center}
\textbf{\large Supplementary Information}
\end{center}
\vspace{2ex}
\twocolumngrid

\setcounter{figure}{0}
\makeatletter 
\renewcommand{\thefigure}{S\@arabic\c@figure}
\renewcommand{\theequation}{S\arabic{equation}}
\renewcommand{\thesection}{S\arabic{section}}
\makeatother

\section{Sample parameters}

	\begin{table}[ht]
\caption {Parameters of the \MZOZO~2DES when $n$~=~2~\density, electron mobility $\mu$~=~600,000~cm$^2$/Vs, $m^*$~=~0.3$m_\mathrm{0}$ and $\epsilon$~=~8.5$\epsilon_0$. }
	{\begin{tabular}{@{}ccccc@{}} \toprule
			Parameter & Magnitude & Units \\
			\colrule
			$k_\mathrm{F}$ & 3.5$\times 10^7$ & m$^{-1}$\\
			$E_\mathrm{F}$ & 1.6$\times 10^{-4}$ & eV\\
			$T_\mathrm{F}$  & 1.85 & K\\
			$\tau_\mathrm{tr}$ & 100 $\times 10^{-12}$ & s\\
			$L$ & 1.5$\times 10^{-6}$ & m\\
			$\rs$ & 27 & \\
			$q_\mathrm{TF}$ & 1.3$\times 10^9$ & m$^{-1}$\\
			$t_\mathrm{WF}$ & 10 $\times 10^{-9}$ & m\\
			\botrule
		\end{tabular}
	}
	\label{Table.parameters}
\end{table}

Table \ref{Table.parameters} collects relevant parameters of the \MZOZO~2DES when $n=2$~\density~and $\mu=600,000$~cm$^2$/Vs. Here, $k_\mathrm{F}$ is the Fermi momentum, $E_\mathrm{F}$ the kinetic energy, $T_\mathrm{F}$ the Fermi temperature, $\tau_\mathrm{tr}$ the transport scattering time, $L$ the mean free path of conduction, $\rs$ the ratio between Coulomb and kinetic energy, $q_\mathrm{TF}$ the Thomas-Fermi screening wave length, and $t_\mathrm{WF}$ an estimate of the wavefunction thickness, based on the data presented in Ref.~\onlinecite{solovyev:2015}. Note that the band effective mass is used in these calculations. The use of the renormalized mass values will be discussed in the subsequent sections. The peak electron mobility of the device is approximately 600,000~cm$^2$/Vs. Ohmic contacts were formed by evaporating Ti (10\,nm) followed by Au (50\,nm) on the sample surface. Indium was additionally soldered upon these pads to improve the contact quality. Qualitatively similar results were obtained both with and without this additional indium layer.

\begin{figure}
	\includegraphics[width=60mm]{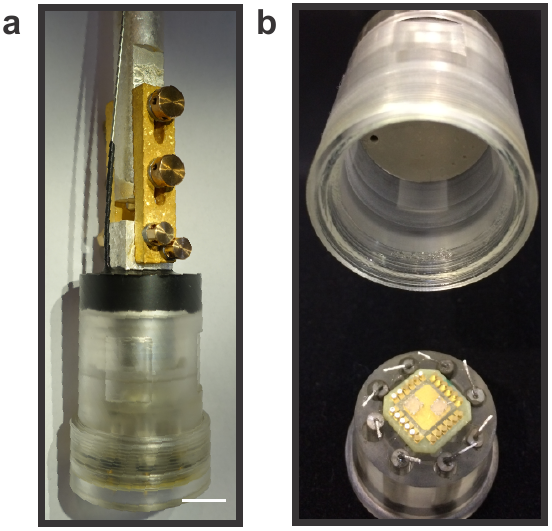}
	\caption{\textbf{The $^3$He immersion cell.} \textbf{a} Closed cell with sample puck inside (scale bar $\approx$1~cm) \textbf{b} Open cell displaying the main heat exchanger and $^3$He inlet hole (top) and sample puck (bottom). The sample is glued upon a metalized chip carrier, with each contact being connected to a sintered silver heat exchanger by a gold bonding wire.}
	\label{S_He3cell}
\end{figure}

\section{Measurement details}

All measurements have taken place in a $^3$He immersion cell anchored at the base of an annealed silver cold finger attached to the mixing chamber of a 650$\Mu$W at 120~\,mK dilution refrigerator (base $T \approx 7$\,mK). The cell design is based on the one utilized at the National High Magnetic Field Laboratory High B/T facility in Gainesville, Florida. This technique has proven powerful in achieving the low electron temperatures required for studies of delicate fractional quantum Hall features,\cite{pan:1999} and also insulating phases in low density 2DES devices.\cite{knighton:2018} The mixing chamber temperature is measured using a calibrated cerous magnesium nitrate paramagnetic thermometer for $T\leq120$~mK, and a ruthenium oxide thermometer for $T\geq50$~mK. These thermometers have good overlap in the range 50$<T<$120~mK. We do not measure the temperature inside the $^3$He cell. As electrons at low temperature are actively cooled primarily via electrical contacts at ultra-low temperatures, the exact electron temperature is dependent on the contact resistance. The Kapitza phonon interfacial resistance ($\propto1/(AT^3)$) may be effectively suppressed by immersing the sample in the cryogen and attaching large surface area ($A$) sintered silver heat exchangers to each measurement wire. Electrical signals are carried from room temperature down to the mixing chamber using individual thermocoax lines, each with a length of approximately 2~m. These lines are thermally anchored at the 1~K, still, 50~mK and mixing chamber plates. These cables act as distributed \textit{R-C} filters and are effective in attenuating stray radiation in cryogenics applications.\cite{zorin:1995} They consist of a resistive stainless steel inner conductor ($R_\mathrm{cable}\approx150$\,$\Omega$) isolated from the outer conductor by a MgO nanoparticle dielectric. The skin effect of the nanoparticles strongly attenuates high frequency signals ($>1$~MHz). An additional \textit{R-C} filter is employed at the mixing chamber. Superconducting Nb-Ti loom is used between the mixing chamber and the measurement puck.

\begin{figure}
	\includegraphics[width=55mm]{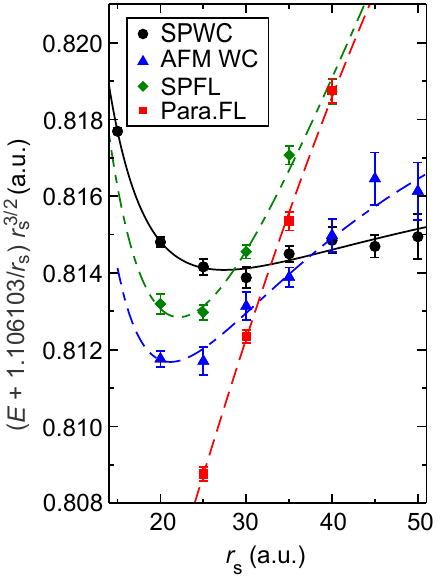}
	\caption{\textbf{Quantum Monte Carlo results adapted from Ref.~\onlinecite{drummond:2009}.} The energy $E$ per electron of competing unpolarized paramagnetic Fermi liquid (Para. FL), spin polarized Fermi liquid (SPFL), unpolarized AFM Wigner crystal (AFM WC) and polarized Wigner crystal (SPWC) as a function of $\rs$.}
	\label{S_drummond}
\end{figure}

\begin{figure*}[t]
	\includegraphics[width=170mm]{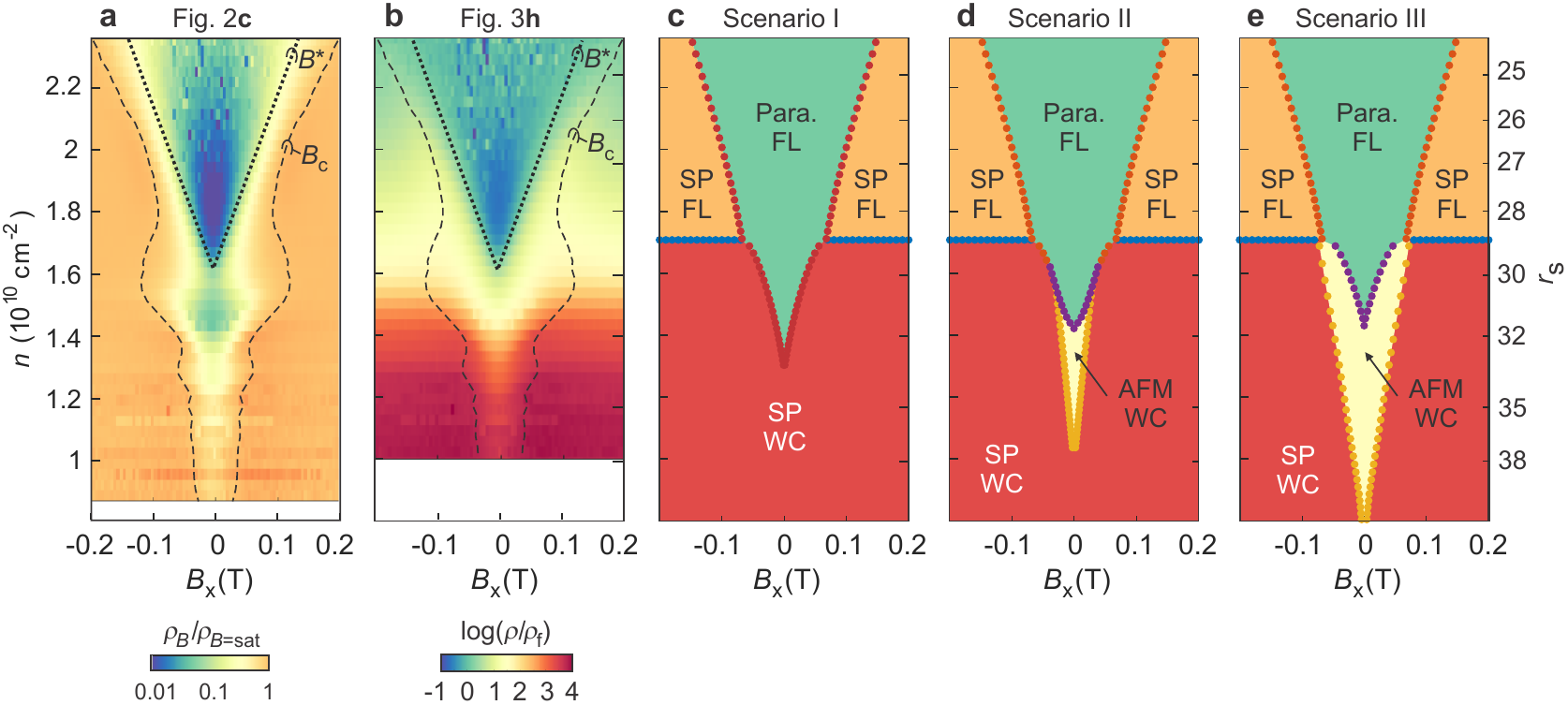}
	\caption{\textbf{Comparison of experimental data with QMC ground states.\cite{drummond:2009}} \textbf{a} $\rho_\mathrm{B}/\rho_\mathrm{B=sat}$ in the ($B_x$,\textit{n})-plane. \textbf{b} Nonlinearity in IV, defined as $\rho/\rho_\mathrm{f}$ in the ($B_x$,\textit{n})-plane. Phases associated with \textbf{c} scenario I, \textbf{d} scenario II and \textbf{e} scenario III, as discussed in Sec.~\ref{sec:SIphasediagram}.}
	\label{S_QMontecarlo}
\end{figure*}

Magnetic flux is generated by a 3-axis (9-3-1 T) superconducting vector magnet. The cryostat is suspended above a pit carved into a 35~ton concrete block that is subsequently isolated from the main building structure by vibration dampening pads. Circulation pumps are anchored to the outer wall of the superstructure to reduce vibrations. The measurement room is electrically isolated from the outside environment by a Faraday cage delivering -60~dBm attenuation. Measurement electronics are powered by a dedicated phase of a three-phase power supply, operate on individual isolating transformers, and are isolated from the data gathering measurement computer by an optical isolator. Significant efforts, such as electrically isolating the cryostat from pumps and diagnostics electronics, have been taken to minimize ground-loops within the measurement circuitry. We strive to utilize low-noise electronics for measurements, such as all analogue lock-in amplifiers (PAR 124A) (only used for the data presented in Fig.\ref{S_BMIT}), Yokogawa 7651 DC voltage sources and DL Instruments voltage (1201) and current (1211) preamplifiers.

DC measurement techniques often incorporate offsets in both the current and voltage signal. These offsets can also fluctuate in time. We have tried to reduce this uncertainty by ensuring excellent temperature stability of the laboratory, while maintaining electrical power to the electronics at all times. The current offset is addressed by adjusting the zero offset of the current preamplifier at zero applied electric field. The voltage offsets are more challenging to eliminate with features order of nano- to microvolts between individual IV sweeps being observed. The only point where we have adjusted the data in this manuscript is in Fig.\ref{Fig4}\textbf{a}-\textbf{c} and \ref{S_twopoint}, where we take the measured voltage at zero current at a set temperature and charge density and subtract it as an offset for each trace. The process of differentiating data eliminates this offset as it is only sensitive to the slope of the data. 

\section{Monte-Carlo Phase Diagrams of 2D Jellium}

The analysis we present here relies on the variational Monte-Carlo study of Ref.~\onlinecite{drummond:2009}. This study compared energies of four states: a spin unpolarized paramagnetic Fermi liquid (Para. FL), spin polarized Fermi liquid (SP FL), spin unpolarized Wigner crystal (AFM WC) with a stripe-like spin antiferromagnetic order and a triangular lattice and a spin polarized Wigner crystal (SP WC) with triangular lattice. The main results of Ref.~\onlinecite{drummond:2009} are adapted in Fig.~\ref{S_drummond}, and imply that as $r_s$ increases, the ground state progresses from Para. FL, to AFM WC, to SP WC, in the vicinity of $\rs=30$. Importantly, Ref.~\onlinecite{drummond:2009} ruled out the possibility of the traditional Stoner transition from Para. FL to a SP FL.

We use this study to estimate the phase diagram as a function $B_\parallel$ and $r_s$. Because electrons in ZnO have weak spin-orbit coupling, the total spin $S_z$ of the system is a good quantum number. This allows us to know the exact energy of each of the four states previously described, which are all either fully spin polarized or fully spin un-polarized. However, to have a complete phase diagram we need to estimate the energy of potential states with different partial polarization. To do this, it is unavoidable to make certain assumptions about the dependence of the energy on spin polarization. We therefore analyze the competition of these phases within three distinct scenarios. As we will see the main conclusions are relatively robust to the assumptions within different scenarios.

\subsection{Scenario I: phase diagrams without a partially polarized crystal}
\label{sec:SIphasediagram}

The appearance of a possible competing intermediate crystal is a relatively new development within the series of Monte-Carlo studies of the ideal Jellium model~\cite{drummond:2009}, although indications of such an intermediate state appeared in a previous study~\cite{bernu:2001}, and also in self-consistent Hartree-Fock calculations~\cite{bernu:2011,bernu:2017}. Since most previous studies did not find or ignored the possibility of these partially polarized crystalline states, it is natural to try to determine what would be the phase diagram if such states are not considered as part of the competing states. To do so we begin by approximating the ground state energy per electron of the Fermi liquids as a function of their spin polarization polarization $p$, given by
\be 
p \equiv \frac{n_{\uparrow }-n_{\downarrow }}{n_{\uparrow }+n_{\downarrow }}, \ p \in [-1,1],
\ee
with a simple parabolic dependence, as follows:

\begin{multline}\label{EFL}
    \epsilon_{FL}\approx \epsilon_{Para.FL}(r_s) \\+p^2 (\epsilon_{SPFL}(r_s)-\epsilon_{Para.FL}(r_s))-\frac{g \mu_B B_x}{2}  p.
\end{multline}
Here $\epsilon_{Para.FL}$ and $\epsilon_{SPFL}$ are the energies of the unpolarized FL and spin polarized FL phases from Ref.~\onlinecite{drummond:2009} and the $g$-factor is $g\approx 2$.\cite{kozuka:2013} Notice from Fig.~\ref{S_drummond} that since $\epsilon_{Para.FL}(r_s)<\epsilon_{SPFL}(r_s)$ for our range of interest $\rs \lesssim 40$, there is always an energy penalty for polarizing the Fermi liquid, and no Stoner transition. In this first scenario, these states compete only with a PWC, whose energy is
\be
\epsilon_{WC}\approx \epsilon_{SPWC}(r_s)-\frac{g \mu_B |B_x|}{2},
\ee
where $\epsilon_{SPWC}$ is the energy of spin polarized WC from Ref.~\onlinecite{drummond:2009}. For each $\rs$ we find the FL state with the optimal polarization and determine its energy competition with the WC. The resulting phase diagram is shown in Fig.~\ref{S_QMontecarlo}\textbf{c}.

\subsection{Scenario II: phase diagrams with a partially polarized crystal with a first order transition}

We now include the possibility of a partially polarized (AFM) WC and model its energy using the same parabolic dependence we used for the FL, as follows:
\begin{multline}
\epsilon_{WC}\approx \epsilon_{AFMWC}(r_s)\\+p^2 (\epsilon_{SPWC}(r_s)-\epsilon_{AFMWC}(r_s))-\frac{g \mu_B B_x}{2}  p
\end{multline}

The phase diagrams obtained by comparing these energies from those of the FL states from Eq.~\eqref{EFL} is shown in Fig.~\ref{S_QMontecarlo}\textbf{d}. Notice that since there is a crossing of the energies of SPWC and AFM WC around $r_s\approx 38$ (see Fig.~\ref{S_drummond}), the above form predicts a first-order spin-flop-type transition of the spin polarization among the WC states at $r_s\approx 38$. There is \textit{a priori} no reason favoring this transition to be first order rather than continuous, which leads us to consider this scenario separately.

\subsection{Scenario III: phase diagrams with a partially polarized crystal with a continuous transition}

In order to account for a possible continuous transition of the spin polarization of the partially polarized Wigner crystal into the fully polarized crystal, we parameterize their energy dependence on spin polarization with a simple quartic Ginzburg-Landau (GL) form:
\begin{multline}
\epsilon_{WC} \approx \epsilon_{AFMWC}(r_s)+ p^2 b(r_s)+ p^4 c(r_s)-\frac{g \mu_B B_x}{2}  p
\end{multline}

As customary for GL quartic functionals, we take the coefficient $c(r_s)$ to be positive for all $r_s$. The crystal starts developing a non-zero spontaneous spin-polarization at a certain density $r_{s1}$, for which the coefficient $b(r_s)$ changes from positive to negative. For $r_{s}>r_{s1}$ the GL functional predicts a monotonically increasing spontaneous polarization, until the system saturates at $p=1$ at some second density $r_{s2}$. With these inputs the coefficients of the GL functional can be fixed to be:
\begin{multline}
b(r_s)=2 (\epsilon_{AFMWC}(r_{s})-\epsilon_{AFMWC}(r_{s1}))\\ \times \frac{\epsilon_{AFMWC}(r_{s2})-\epsilon_{SPWC}(r_{s2})}{\epsilon_{AFMWC}(r_{s2})-\epsilon_{SPWC}(r_{s1})},
\end{multline}

\be
c(r_s)=\epsilon_{SPWC}(r_{s})-\epsilon_{AFMWC}(r_{s})-b(r_s).
\ee

Since the Monte Carlo study \cite{drummond:2009} did not explore partially polarized states, the precise values of  $r_{s1}$ and $r_{s2}$ are unknown, although they should satisfy  $r_{s1}<38<r_{s2}$. The phase diagram shown in Fig.~\ref{S_QMontecarlo}\textbf{e} has been made by choosing $r_{s1} = 27$ and $r_{s2} = 45$. The precise of form of the phase diagram does not have strong sensitivity on the precise values of $r_{s1}$ and $r_{s2}$, provided they are chosen in the vicinity of $r_{s}\sim 38$ up to some distance of about $\Delta r_s \sim 10$. In fact, in the limiting case in which the polarization changes rapidly from unpolarized to polarized, $r_{s1}$ approaches $r_{s2}$ and one recovers the Scenario II of a first order transition. The main difference is a reduction in size of the region where the partially polarized crystal is energetically favored, but the overall shape of the two phase diagrams still looks reasonably similar in both scenarios II and III, shown in Fig.~\ref{S_QMontecarlo}\textbf{d} and \textbf{e}.

\begin{figure*}
	\includegraphics[width=170mm]{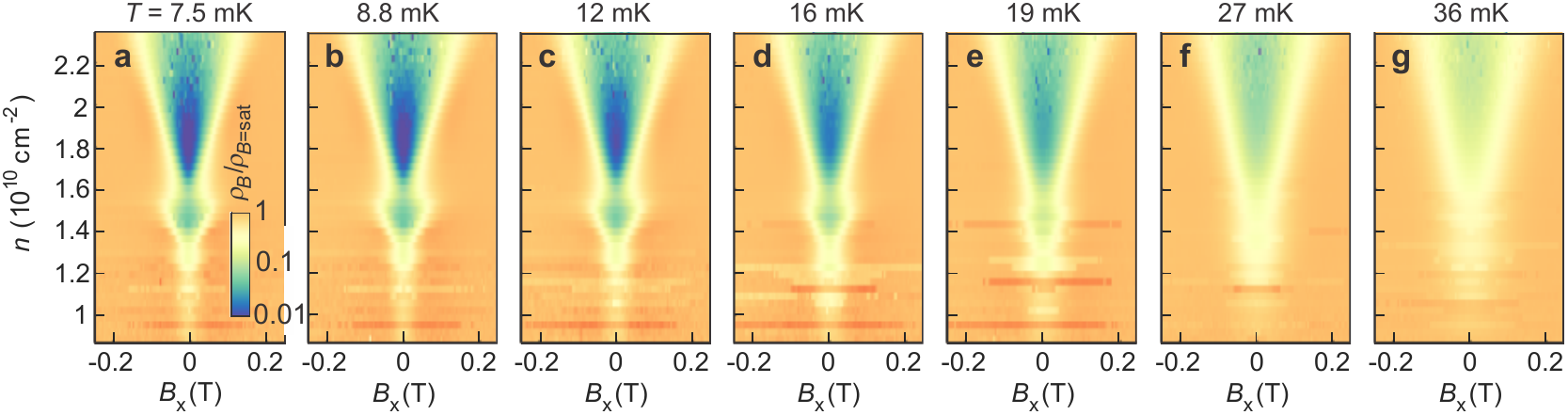}
	\caption{\textbf{Temperature dependence of in-plane magnetic field mapping.} \textbf{a-g} Mapping of the normalized differential resistance $\rho/\rho_\mathrm{B=sat}$ as a function of $\Bx$~and $n$ at different temperatures.}
	\label{S_TdepBx}
\end{figure*}

\section{Landau Fermi liquid parameters}
\label{sec:SIFLparameters}

The energy density of the metallic system at fixed density in the presence of an in-plane Zeeman field, $B_x$,  is given by
\begin{equation}
E\approx E_0+\frac{s_z^2}{2 \chi} - g_0 \frac{\mu_B}{\hbar} s_z B_x + \mathcal{O}(s_z^4),
\end{equation}
where $s_z=(\hbar/2)(n_\uparrow-n_\downarrow)$ is the spin density, $\chi=m^*/2\pi$ the spin susceptibility, $g_0$ is the $g$-factor, and $\mu_B$ is the Bohr magneton. Therefore, the critical field, $B_c$, at which the system fully spin polarizes at a given total density, $s^{\rm pol}_z=(\hbar/2) n$, provides an approximate measure of the spin susceptibility of the system:
\begin{equation}
B_c \approx \frac{\hbar^2 n}{2 g_0\mu_B \chi}.
\label{eq:SIBc}
\end{equation}

In Landau Fermi liquid theory, this susceptibility is expressed in terms of the quasiparticle mass ($m^*$), the bare band mass ($m_b$), the bare spin susceptibility ($\chi_0$) and the $F_s(0)$ Landau parameter as follows~\cite{pines:2018}:
\begin{equation}
\chi =  \chi_0 \frac{1}{1+F_s(0)}\frac{m^*}{m_b}.
\end{equation}

\begin{figure}
	\includegraphics[width=65mm]{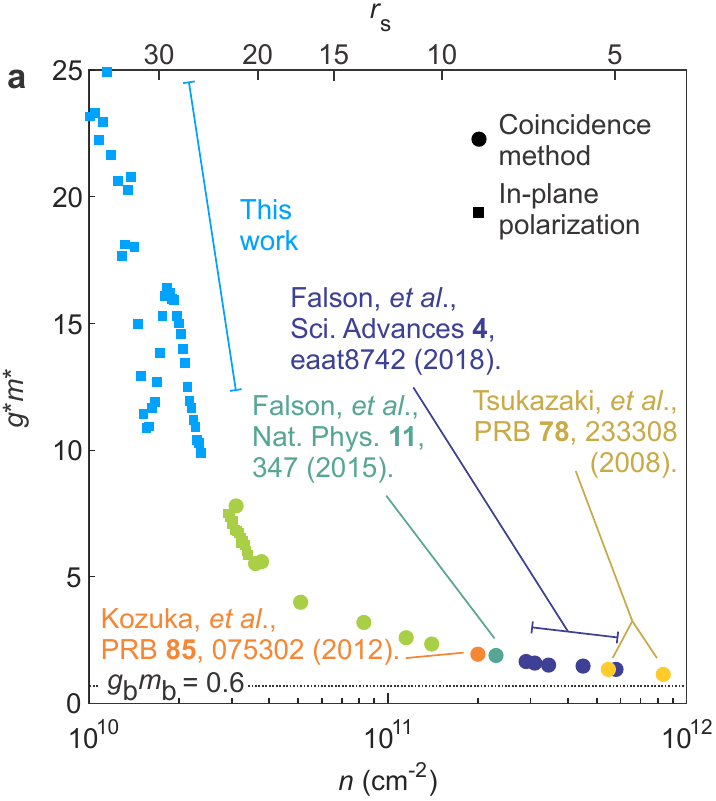}
	\caption{\textbf{Summary of renormalization effects in ZnO devices.} $g^*m^*$ as a function of $n$. The band value $g_\mathrm{b}m_\mathrm{b}$=0.6 is displayed as a horizontal dashed line. Data have been gathered from devices analyzed in this text, previous publications, and other unpublished measurements.\cite{tsukazaki:2008,kozuka:2012a,falson:2015a,falson:2018b}}
	\label{S_gm}
\end{figure}

\begin{figure}
	\includegraphics[width=56mm]{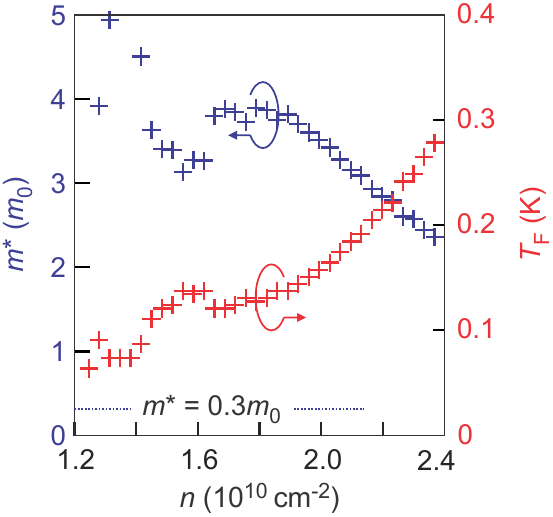}
	\caption{Renormalized effective mass and Fermi temperature as a function of $n$.}
	\label{S_effectiveemass}
\end{figure}

The spin susceptibility of carriers can be evaluated by various means. A strategy we have employed in previous studies is the coincidence method which relies on rotating the sample within a magnetic field to study transport features associated with the crossing of opposing spin Landau levels.\cite{tsukazaki:2008,falson:2015a,falson:2018b} In addition to the coincidence method, it is possible to polarize carriers into a single spin band by applying an in-plane magnetic field. According to Eq.\ref{eq:SIBc}, identification of a critical field $\Bc$ leads to the quantification of $g^*m^*$.  This measurement is discussed in the main text in the context of Fig.~\ref{Fig3} with saturation of the magnetoresistance at a critical field $\Bc$ being taken as evidence for full spin polarization. For the sake of completeness, here we present the full raw data set used to produce the analysis presented in Fig.~\ref{Fig2} of the main text. This data set comprises multiple maps of the differential resistance in the ($n$,$\Bx$)-parameter space at a number of set temperatures. These are all presented in Fig.~\ref{S_TdepBx}. As in Fig.~\ref{Fig2}\textbf{c}, we plot the ratio $\rho/\rho_\mathrm{B=sat}$, where $\rho_\mathrm{B=sat}$ is the saturated differential resistance at a set charge density at high magnetic field (corresponding to full spin polarization). The characteristic positive magnetoresistance is evident for all $T$ at high $n$, but is difficult to discern in the higher $T>20$\,mK data upon depletion of the 2DES.

The renormalization of band parameters in the ZnO-based 2DES as a function of charge density is summarized in Fig.~\ref{S_gm}. This figure is compiled from values reported in previous publications\cite{kozuka:2012a,falson:2015a,falson:2018b}, the devices presented in this manuscript, along with other unpublished results gathered on samples characterized in the course of this project. Values obtained through the coincidence method are displayed as circles, with squares representing those gained through analyzing the positive magnetoresistance of a polarizing in-plane magnetic field. The error associated with each point is no larger than the symbol size. A good agreement between these two methods is obtained for the range of densities shown. We additionally display the corresponding $\rs$ value at each $n$ on the top axis. Previous studies on Si-based devices identified that the enhancement of the spin susceptibility as the critical density is approached is associated with an enhancement of the effective mass.\cite{shashkin:2002}

Utilizing the experimentally obtained $g^*m^*$ and working under the assumption that the parameter enhancement is primarily due to the renormalization of the effective mass,\cite{falson:2018a} we can extract the mass value and renormalized Fermi temperature ($T^*_\mathrm{F}$) as a function of $n$. This is plotted in Fig.~\ref{S_effectiveemass}. According to Ref.~[\onlinecite{lilly:2003}], a local ``peak'' in $\partial r/\partial T|_n$ of dilute GaAs-based 2DES emerges as $T$ becomes comparable but smaller than $T_\mathrm{F}$ given $T<T_\mathrm{BG}$, where  $T_\mathrm{BG}$=2$k_\mathrm{F}\hbar v_\mathrm{s}/k_\mathrm{b} \approx$ 2.4~K is the Bloch-Gr\"uneissen temperature calculated using $v_\mathrm{s}=4400$~m/s as the average between transverse and longitudinal sound velocities in ZnO and $n=2$\density. Being deep in the Bloch-Gr\"uneissen regime, phonon scattering contributes little to the resistivity. A local peak in $\partial \rho/\partial T|_n$ is an experimental feature is observed in our data for {$n>1.6$\density}. The calculated $T^*_\mathrm{F}$ values are those presented in Fig.\ref{S_Egap}. The local maximum in $\rho(T)$ when $n>\nc$ occurs just below the calculated $T^*_\mathrm{F}$. We note that the corresponding calculated Fermi temperature is in the range $T_\mathrm{F}=1.5\sim1.7$~K utilizing the band mass. 

\begin{figure}[h]
	\includegraphics[width=85mm]{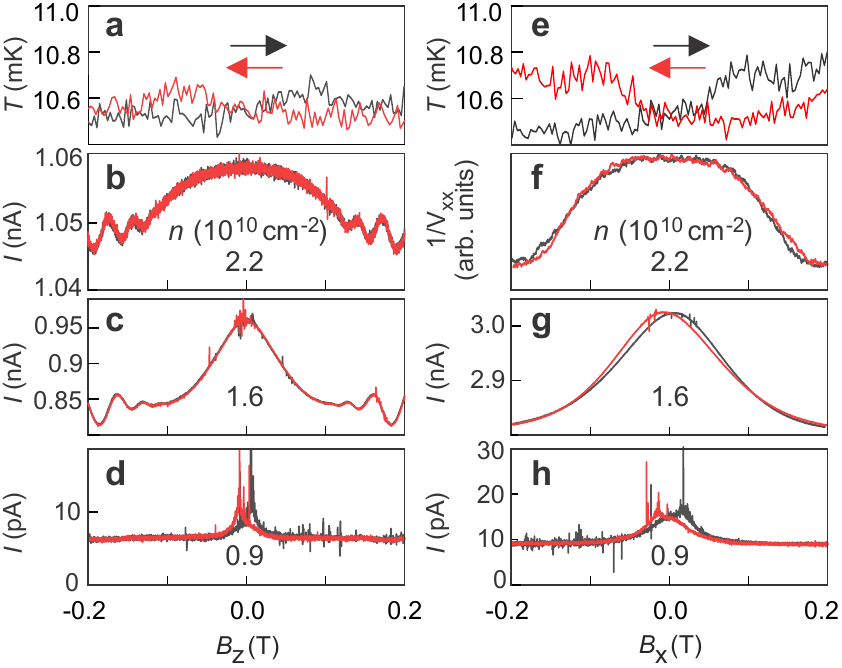}
	\caption{\textbf{Sweep direction dependent magnetotransport in the vector magnet equipped cryostat.} \textbf{a} Temperature read at the mixing chamber plate as a function of sweep dependence for an out-of-plane magnetic field ($B_\mathrm{z}$). \textbf{b-d} Measured electrical characteristics for three different charge densities. Panels \textbf{e-h} plot analogous data with the magnetic field projected in the in-plane direction ($B_\mathrm{x}$).}
	\label{S_hysteresis1}
\end{figure}

\section{Absence of magnetostatically induced hysteresis and domains in ferromagnetic phases}
\label{sec:S_hysteresisdiscussion}

Here we estimate the typical value of the coercive fields associated with magnetic hysteresis arising from magneto-static fields. Our goal is to argue that these fields are too small to be responsible for any observed behavior in our samples. We clarify that our argument does not imply that the system cannot, in principle, display hysteresis via other mechanisms, such as metastability associated with pinning of the Wigner crystal. But we are able to argue that the non-trivial in-plane field dependence of the resistivity of the insulating state, seen in Fig.\ref{Fig3}\textbf{b} of main text, cannot be a result of magnetic-field-induced ferromagnetic domain alignment of the putative ferromagnetic Wigner crystal.

Since ZnO has negligible spin-orbit coupling, the formation of magnetic domains and hysteresis is dictated by magneto-statics. Assuming that each electron contributes one Bohr magneton, $\mu_B$, to the magnetization in the spin polarized state, the coercive field scale (the magnetic field width of hysteresis loop) can be estimated to be:
\be
B_c = \mu_0 \mu_B \frac{1}{V_{\rm elec}}=\mu_0 \mu_B \frac{ n}{d},
\ee
where $\mu_0$ is the vacuum magnetic permeability constant, $V_{\rm elec}=d/n$ is the average volume per electron, and $d\approx 5$\,nm is the effective well width of the 2DES in ZnO. For $n\sim 10^{10} cm^{-2}$ we have:

\be
B_c  \sim 2.4 \times 10^{-7} {\rm T}.
\ee

The smallness of this scale is ultimately a consequence of the tremendous diluteness of our systems. Each electron occupies a 3D volume of about $V_{\rm elec} \sim 5\times 10^7 $  \AA$^3$, which is about $5 \times 10^6$ larger than in ordinary 3D metals such as copper. The above strongly indicates that magnetostatic domain formation is completely negligible.

We now present our experimental data close to zero field for the sake of completeness. After much effort to properly understand these effects, we emphasize that it is extremely difficult to eliminate all parasitic effects when collecting the data. There are a number of experimental unknowns that are difficult to avoid. Firstly we mention temperature fluctuations when sweeping the magnetic field due to the presence of nuclear and electron spin magnetization processes. These may occur in the sample,  electrical contacts, experimental wiring, and materials used to construct the helium immersion cell. Furthermore, it is very difficult to exclude the influence of remnant magnetic flux in the superconducting magnet system used. 

Figure~\ref{S_hysteresis1} presents sweep-direction-dependent electrical characteristics of the device measured. The sample is biased with a DC voltage that produces approximately 1~nA when charge is accumulated and the sample is metallic. Panels \textbf{a-d} present data when the field is projected in the $z$ direction, with \textbf{e-g} presenting data in the $x$ direction.  We measure the thermometer reading at the mixing chamber through these sweeps, which are performed at a very low rate of 1~mT/min, as shown in panels \textbf{a} and \textbf{e}. Sweeping $\Bz$ conveniently reveals quantum oscillations, whose period are known to be determined by the magnetic flux penetrating the sample, and therefore may be used to reduce the uncertainty of remnant flux in the coil at low field by forcing the oscillation minima in up-sweep and down-sweep data to overlap. Panels \textbf{b-d} and \textbf{f-h} present sweeps at three distinct charge densities, corresponding to the metallic, critical and insulating portions of the parameter space. Within our detection limit, the up- and down-sweep data overlap closely, as shown in panel \textbf{b}. Operating under the assumption that hysteresis is weak or absent at this charge density, we enforce up- and down-sweep data to overlap measured in the $\Bx$ direction, as shown in panel \textbf{f}. This process establishes a remnant flux of approximately 5~mT in the $\Bz$ and 15~mT in the $\Bx$ coils. This value is used to manually offset subsequent field sweeps performed in a repetitive cycle in an attempt to reduce the impact of this parasitic effect. Next, we can vary the charge density of the device, as plotted in panels \textbf{c} and \textbf{g} when $n$ is close to $\nc$. While higher field quantum oscillations overlap nicely in the $\Bz$ data set, additional delicate features close to $B=0$ in the form of sharp spikes in the data are resolved in both projections of the magnetic field. These become even more evident upon reducing $n$ below $\nc$. While these features occur at fields where the temperature read during up and down sweeps coincides, it is not possible for us to know what is occurring locally at the sample. These features do not follow a Curie-Weiss-like temperature dependence and are robust only when the $|d\rho/dT|$ of the device is high. Hence, in the absence of direct spin-sensitive probes and an estimate of the coercive field, we associate the hysteresis with an uncontrollable experimental aspect of the measurement.

\section{Exchange scale estimate and impact of temperature fluctuations on spin order}\label{sec:S_Xchangediscussion}

\begin{figure}
\begin{center}
\includegraphics[width=50mm]{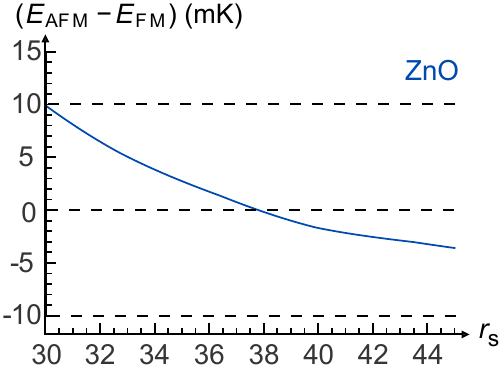}
\caption{ Energy difference per electron of the AFM WC and FM WC from QMC~\cite{drummond:2009}, with parameters adjusted for ZnO in units of mK.}
\label{ZnOXchange}
\end{center}
\end{figure}

To estimate the typical scale of the spin exchange energy of the Wigner crystal, $J$, we will compare the energy difference per electron of the AFM WC and the FM WC obtained from QMC in Ref.~\onlinecite{drummond:2009}. In Fig.~\ref{ZnOXchange} we plot this energy difference in units of mK, by using parameters relevant for ZnO, given by $m\approx 0.29 m_0$, $\epsilon\approx 8.5$, as a function of $r_s$. We also display horizontal dashed lines corresponding to a temperature scale $T=10$mK. We see that the energy difference never exceeds this temperature scale over the entire range of $r_s\in [30,45]$, and in particular, it is clearly smaller than this scale in the Ferromagnetic side. This suggests that our experimental temperatures are at best only comparable with the scale needed to destroy the spin ordering of the WC at $B_x=0$.

\begin{figure}
\begin{center}
\includegraphics[width=50mm]{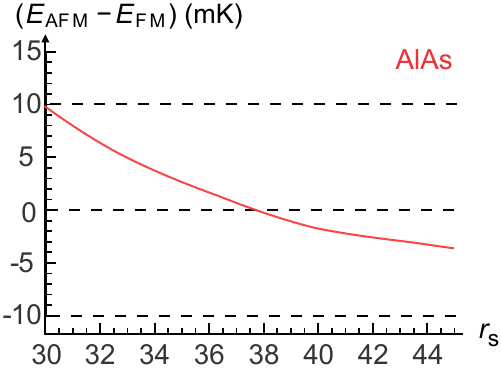}
\caption{Energy difference per electron of the AFM WC and FM WC from QMC~\cite{drummond:2009}, with parameters adjusted for AlAs in units of mK.}
\label{AlAsXchange}
\end{center}
\end{figure}

For comparison we plot the same energy difference converted to the scales relevant to AlAs two-dimensional electron systems studied in Ref.~\onlinecite{hossain:2020}, by taking $m\approx 0.4 m_0$, $\epsilon\approx 10$ in Fig.~\ref{AlAsXchange}. The value $m \approx 0.4 m_0$ is the geometric mean of the two orthogonal masses. We note that one difference between ZnO and AlAs is the presence of an anisotropic dispersion and multiple valleys of carriers in the latter. By chance the effective unit of energy, $\hbar^2/(m a^2)$, is nearly identical in both settings, ZnO and AlAs, with $\hbar^2/(m a^2)  \approx 1260$\,K. (Here $a$ denotes the effective Bohr radius.) For this reason Fig.~\ref{ZnOXchange} and Fig.~\ref{AlAsXchange} look almost identical.

Since the lowest temperature reached in Ref.\onlinecite{hossain:2020} was $T\approx 300$mK, it is unlikely that the system reaches the regime of ferromagnetic ordering of the spins of the WC. An alternative explanation for their observations at lowest densities is that the flatness of the magnetoresistance trace as a function of $B_x$ might no longer be a reliable criterion for establishing spin polarization deep in the insulating regime at such elevated temperatures. As we see in our case, even a relatively modest increase of the temperature to about $T = 40$\,mK, is enough to flatten out the magnetoresistance as a function of $B_x$ at the lowest densities (see e.g. Fig.\ref{Fig3}\textbf{c} of main text and Fig.\ref{S_TdepBx}).


\section{Temperature dependence of the resistance}
\label{sec:SItemperaturedependence}

\begin{figure}[ht]
	\includegraphics[width=85mm]{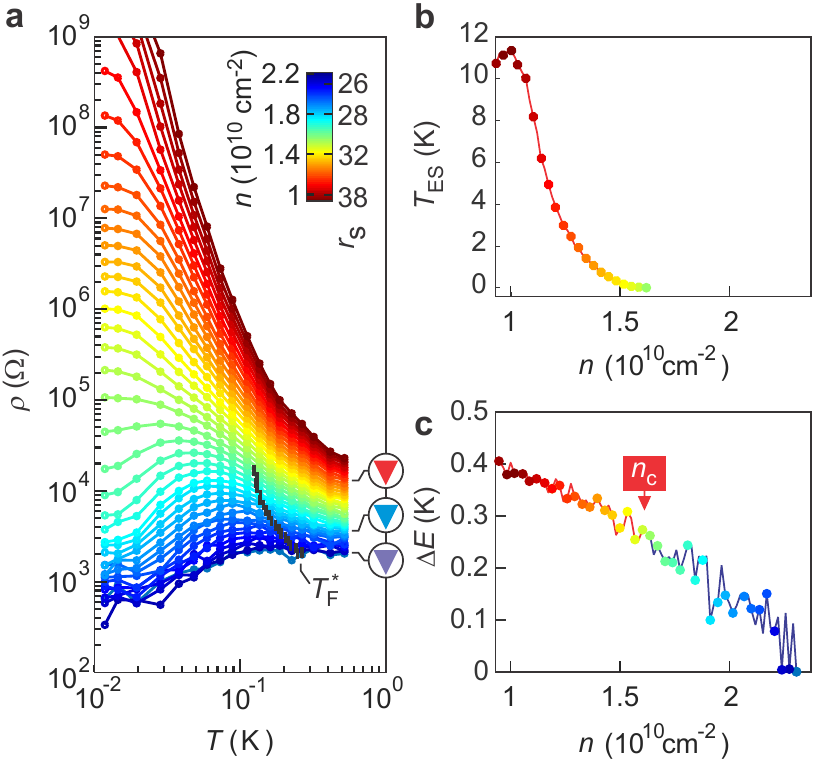}
	\caption{\textbf{Temperature dependent transfer characteristics.} \textbf{a} MIT in the temperature dependence of the resistance as $n$ is tuned. \textbf{b} $T_\mathrm{ES}$ as a function of $n$. \textbf{c} Estimated activation energy $\Delta E$ as a function of electron density $n$, obtained by fitting the temperature range $300$\,mK $ < T < 700$~mK.}
	\label{S_Egap}
\end{figure}

We divide the data presented in Fig.~\ref{S_Egap}\textbf{a} into separate regimes of temperature in order to perform an analysis of the temperature dependence of the resistance. In this section we first discuss the behavior at zero magnetic field before turning to the behavior at $B_x > B_c$.

\begin{figure*}[ht]
	\includegraphics[width=155mm]{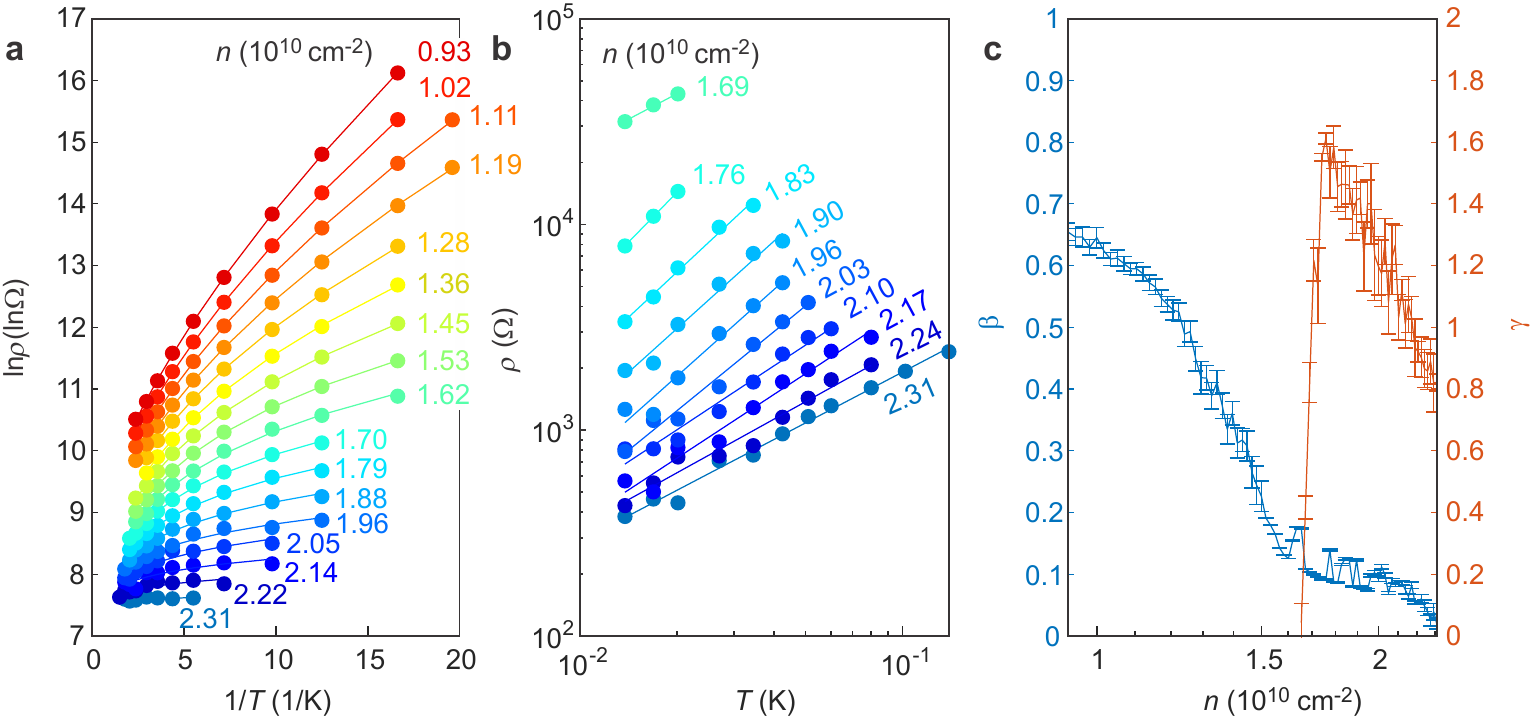}
	\caption{Temperature scaling analysis in \textbf{a} the insulating and \textbf{b} the metallic regime. \textbf{c} Summary of temperature scaling exponents.}
	\label{S_tempscaling}
\end{figure*}

\subsection{Zero magnetic field}

In electronic insulating phases, the temperature-dependent resistivity is generically described by
\begin{equation}
\rho(T) \propto \frac{1}{T^\alpha} \mathrm{exp} \left[ \left( \frac{T_0}{T} \right)^\beta \right].
\label{Eq.insulatingrho}
\end{equation} 
At low temperatures the exponent $\beta$ dominates the behavior of the resistance, and it can take a range of values depending on the mechanism for transport.  These include  $\beta = 1$ for Arrhenius-type activated transport, $\beta = 1/2$ for Coulomb-gap mediated variable range hopping, and $\beta = 1/3$ for Mott-like variable range hopping.\cite{shklovskii_electronic_1984} A common method for estimating the exponent $\beta$ is to perform a Zabrodskii-Zinov'eva analysis, \cite{zabrodskii_low-temperature_1984} in which one defines the (dimensionless) reduced activation energy
\be 
W(T) = - \frac{d \ln \rho}{d \ln T} = \beta \left( \frac{T_0}{T} \right)^\beta + \alpha.
\ee 
The analysis in Fig.~\ref{S_Egap}\textbf{b} suggests that at $n < n_c$ there exists a regime of $\beta = 1/2$ at low temperature that crosses over to a regime of $\beta = 1$ at higher temperature, as in conventional semiconductors on the insulating side of the doping-induced metal-insulator transition \cite{shklovskii_electronic_1984}. (It is particularly difficult to extract reliable estimates of $W(T)$ at $T \lesssim 20$\,mK due to possible decoupling of the electron temperature from that of the cryogen.) Within the regime of $\beta = 1/2$, the value of the temperature $T_0$ (which can be called the ``Efros-Shklovskii temperature'' $T_\textrm{ES}$) depends on the localization length $\xi$ as $T_\textrm{ES} = 6.2 e^2/(4 \pi \epsilon k_B \xi)$ \cite{shklovskii_electronic_1984}.  We observe that $T_\textrm{ES}$ vanishes as the electron density approaches $n = n_c$ from below (Fig.~\ref{S_Egap}\textbf{c}), which is consistent with a divergence of the localization length at the metal-insulator transition.

An alternative and more direct analysis for assessing the temperature dependence of potentially insulating states is to perform a fit to the form of Eq.~\ref{Eq.insulatingrho} over all regimes of temperature and density for which $d\rho/dT < 0$ (the usual definition of ``insulating-like'' temperature dependence). Upon close inspection, such a regime appears for all $n$; it is weak but notable even at our maximum value of $n$ under the condition $T>T^*_\mathrm{F}$. Figure \ref{S_tempscaling}\textbf{a} plots $\ln \rho$ as a function of $1/T$ at a number of $n$ at zero magnetic field. The solid lines represent individual fits at discrete $n$. For this process, we avoid using the lowest temperature data points in the strongly insulating regime owing to our inability to exactly determine the electron temperature as the base temperature $T$ is approached, and also due to experimental difficulties associated with determining $dV_\mathrm{xx}/dI$ at very low currents in strongly non-linear $I-V$ traces. Figure \ref{S_tempscaling}\textbf{c} plots $\beta$ as a function of $n$. The exponent $\beta$ is small (insulating-like behavior is weak) at $n>\nc$, but rises rapidly as $\nc$ is reduced below $\nc$. The value of $\beta$ ultimately appears to saturate at a value of approximately $\beta \approx 0.65$ at the lowest $n$.

We separately examine the metallic regime $n<\nc$, for which we fit the data according to a power law,
\begin{equation}
\rho(T) \propto T^\gamma.
\label{Eq.metallicrho}
\end{equation} 
For this analysis, we restrict the range of temperatures to approximately 16 mk $<T<0.8 T_\mathrm{F}|_n$. This is the temperature scaling of the degenerate liquid when $T<T^*_\mathrm{F}$.  We exclude the lowest temperature data points as we are unable to verify that the electron temperature of the sample is accurately reflected by the thermometer based at the mixing chamber at very low temperatures. Figure \ref{S_tempscaling}\textbf{b} presents a log-log plot of the data illustrating a gradually changing slope (\textit{i.e.}~$\gamma$) as $n$ is tuned. The value obtained is plotted as red in Fig.~\ref{S_tempscaling}\textbf{c}. It can be gauged that $\rho(T)$ is approximately linear in $T$ ($\gamma$=1) when $n\approx 2$\density, with $\gamma$ trending towards a value of 1.6 with decreasing $n$ immediately before the onset of the MIT. At lower $n$ metallicity is rapidly suppressed, and the insulating behavior becomes prominent for all $T$.

\subsection{Evolution with in-plane field}

\begin{figure}[ht]
	\includegraphics[width=85mm]{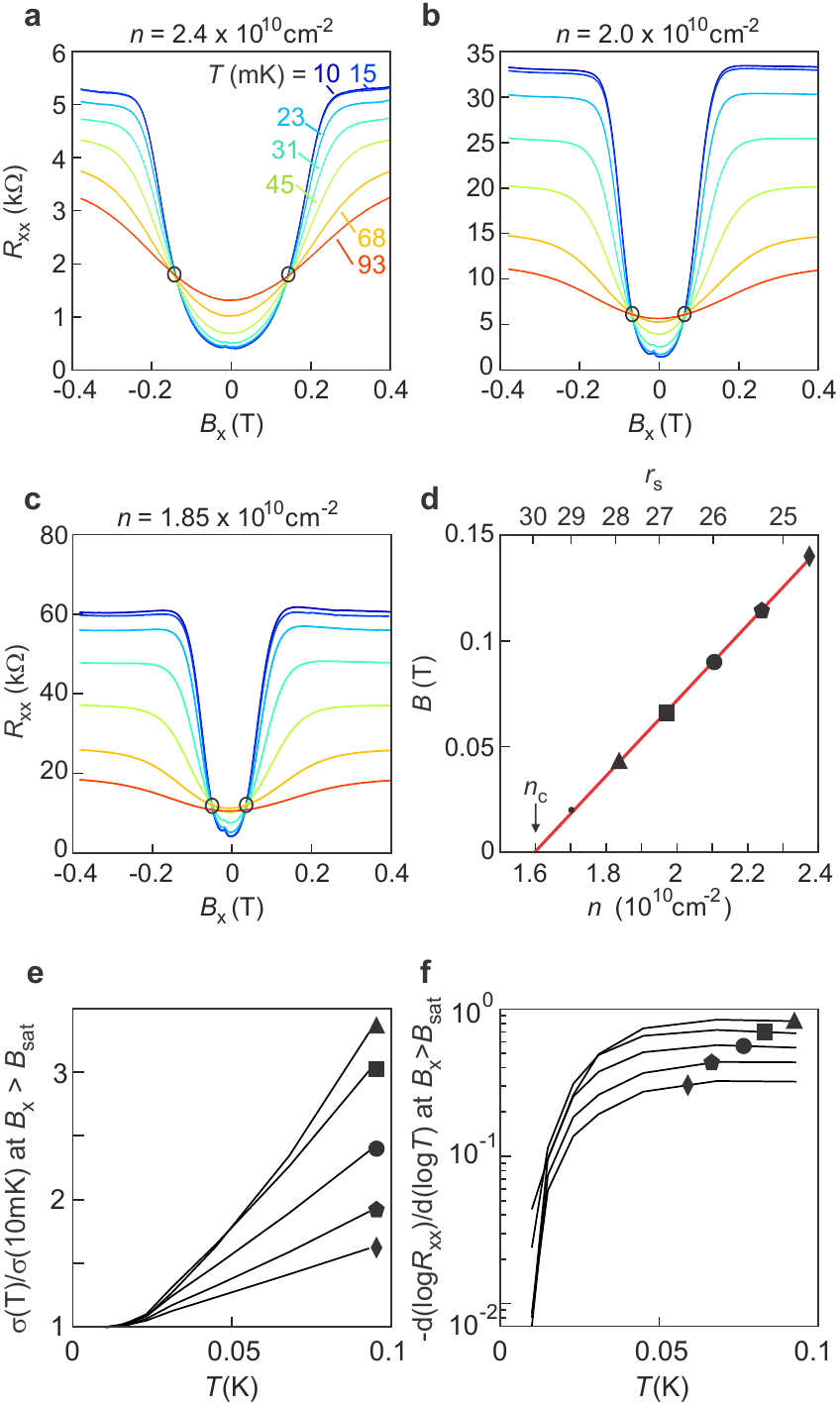}
	\caption{\textbf{Estimate of the critical carrier density $\nc$ through temperature dependent in-plane magnetic field sweeps.} Panels \textbf{a}-\textbf{c} AC magnetotransport sweeps ($I_\mathrm{AC}=2$~nA) recorded  from positive to negative polarity field. The colors correspond to temperatures noted in panel \textbf{a}. Open circles correspond to the field-induced change in temperature dependence. \textbf{d} The identified magnetic field for the change in temperature dependence as a function of $n$. A linear regression through the experimental points (black dots) results in an estimation of the critical carrier density for the zero field MIT occurring at $\nc = 1.6$\density. \textbf{e} Temperature dependence of the normalized conducitivity for charge densities studied in panel \textbf{d}. \textbf{f} Zabrodskii-Zinov'eva analysis.\cite{zabrodskii_low-temperature_1984}}
	\label{S_BMIT}
\end{figure}

As discussed above, applying an in-plane magnetic field acts to polarize the 2DES into a single spin band. Previous measurements in silicon-based devices have suggested that in-plane field also produces an apparent field-induced MIT at an intermediate partial spin polarization.\cite{shashkin:2001} We examine this effect in Fig.~\ref{S_BMIT} by plotting the longitudinal resistance as a function of $\Bx$~at various $T$. Here, we present AC measurements recorded at a current of 2~nA. Panels \textbf{a-c} present the raw magnetotransport data at three distinct $n$. It is possible to identify a change in the sign of $d\rho/dT$ at a finite field $B^{*}$,
at which all the temperature curves coincide. The value of $B^{*}$ is identified as an open circle in Fig.~\ref{S_BMIT} at both polarities of the field. Tracking the value of $B^{*}$ as a function of $n$ gives the results plotted in panel \textbf{d}. Extrapolating this value to $B^{*} = 0$ gives a density 1.6\density~at zero field, which coincides with the value of $\nc$ obtained by DC measurements and discussed in the main text.

In the regime of larger-than-critical density, $n > n_c$, and large in-plane field, $B > B_c$, the 2DES exhibits insulating-like temperature dependence $d \rho / dT$. However, the form of the temperature dependence in this regime is somewhat weaker than in the regime of $n < n_c$ discussed above.  This weak dependence can be seen in Fig.~\ref{S_BMIT}\textbf{e}, which shows that the conductivity $\sigma$ increases roughly linearly with temperature (rather than exponentially). 
Such a linear dependence is qualitatively consistent with a perturbative calculation by Zala et.~al.~\cite{Zala2001rapid, Zala2001interaction}, who found that the conductivity of a spin-polarized FL has a positive, linear-in-$T$ correction due to electron-electron interactions.
The reduced activation energy $W(T)$ in this regime (and at $T \gtrsim 20$\,mK) is also consistent with $W(T) = \textrm{const.}$ (Fig.~\ref{S_BMIT}\textbf{f}).

\section{Disorder-induced density modulation}
\label{sec:SIdisorder}

The phases of the 2DES are generally discussed in terms of a spatially uniform electron density $n$. But when the 2DES is subjected to disorder in the form of stray electric charges, these charges create a random electric potential that modulates the density of the 2DEG spatially.  As mentioned in the main text, this modulation allows for the possibility of multiple phases coexisting in different regions of the sample.

A typical source of disorder for a 2DES is a finite concentration of uncontrolled impurity charges embedded three-dimensionally in the substrate. Due to the long-ranged nature of the Coulomb interaction, such impurity charges produce an arbitrarily large variation in the electrostatic potential over sufficiently long length scales, even when their bulk concentration $N_\textrm{imp}$ is very small, unless they are screened by the 2DES itself. So long as $N_\textrm{imp}$ is sufficiently small (discussed below), this screening can be described using the Thomas-Fermi approximation. \cite{ando_electronic_1982} The corresponding root-mean-square variation $(\delta n)^2$ in electron concentration is given by\cite{skinner_theory_2013} 
\be 
\delta n = \sqrt{ \frac{e^4 N_\textrm{imp}}{4 \pi \epsilon} \left| \frac{d n}{d \mu} \right| }.
\ee 
The quantity $dn/d\mu$ represents the thermodynamic density of states ($\mu$ denotes the chemical potential), and can be either positive or negative, depending on the strength of electron correlations. \cite{Bello_Levin_Shklovskii_Efros_1981, eisenstein:1992}

For the strongly-interacting states that we are considering, $dn/d\mu$ is of order $4 \pi \epsilon n^{1/2}/e^2$ in magnitude, and consequently 
\be 
\delta n \sim (N_\textrm{imp}^2 n)^{1/4}.
\ee 
For our MgZnO/ZnO heterostructures, the bulk impurity concentration is estimated to be below $10^{14}$\,cm$^{-3}$, \cite{li:2013} so that the corresponding density modulation satisfies
\be 
\delta n \lesssim 0.3 \times 10^{10} \textrm{ cm}^{-2}.
\ee 

The Thomas-Fermi approximation that we have employed in this section is valid so long as the typical distance between impurity charges, $N_\textrm{imp}^{-1/3}$, is much longer than the Thomas-Fermi screening radius $(2 \epsilon/e^2)(d\mu/dn)$, which in our case amounts to $N_\textrm{imp} \ll n^{3/2}$. This inequality is easily satisfied in our devices.

\section{Heating}\label{heating}

It is difficult to quantify the extent to which heating influences electrical transport characteristics at very low temperatures. Heating can appear through both current-induced Joule heating (proportional to $I^2R$ in an Ohmic system), as well as magnetic processes as the polarity of the applied magnetic field is changed. From our experiences in higher charge density devices ($n= 2\sim5 \times 10^{11}$~cm$^{-2}$) with similar geometries, such as those studied in Refs.~[\onlinecite{falson:2015a,falson:2018b,maryenko:2018}], Joule heating is negligible in the analysis of fractional quantum Hall energy gaps for currents below approximately 10~nA. However, the charge density, resistivity, and proximity to the MIT of the device under study here puts it in a different parameter space, necessitating a reevaluation of the role of current-induced heating. 

\begin{figure}[h]
	\includegraphics[width=85mm]{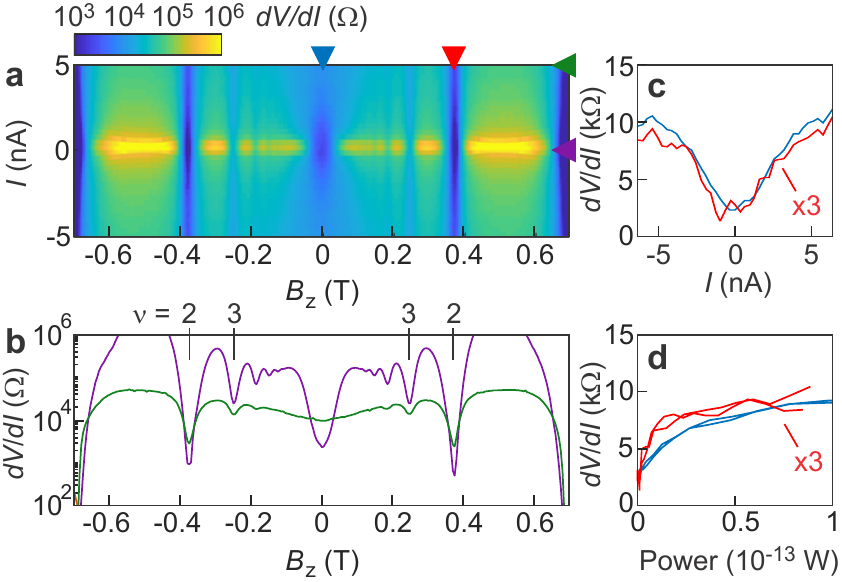}
	\caption{\textbf{Examination of the role of current heating recorded at base temperature of the cryostat. Here, $n$=1.8\density.} \textbf{a} Mapping of magnetotransport in the ($\Bz$,$I$)-plane. \textbf{b} Line cut of the transport taken at $I$=0~nA (purple) and 5~nA (green). \textbf{c} $dV/dI$ as a function of $I$ at $\Bz$=0~T (blue) and 0.375~T (red, corresponding to $\nu$=2). \textbf{d} $dV/dI$ as a function of dissipated power ($P=IV$), also recorded at $\Bz$=0~T (blue) and 0.375~T (red).}
	\label{S_heating}
\end{figure}

Figure \ref{S_heating} examines the role of current-induced Joule heating by examining transport in the ($\Bz$,$I$)-plane. Panel \textbf{a} renders the differential resistance as a function of perpendicular field and current, with two line cuts being presented in panel \textbf{b}. Integer quantum Hall features are visible as vertical deep blue lines in panel \textbf{a}, and deep minima in panel \textbf{b}. The U-shaped magnetoresistance is evident in the purple trace ($I\rightarrow0$~nA), but less so in the finite current trace (green, $I\approx5$~nA). Conventionally, quantum Hall features exhibit an activated temperature dependence ($\rho \propto \mathrm{exp}(\Delta_\mu/T)$), where $\Delta_\mu$ is the activation gap of the state. The depth of the minimum is strongly dependent on temperature, becoming deeper as $T$ is reduced. We leverage this character to evaluate the role of current-induced heating. Two traces of $\rho$ as a function of $I$ are presented in panel \textbf{c}. The blue trace corresponds to $\Bz=0$~T, with the red trace taken at the $\nu=2$ minimum $\Bz=0.375$~T). $n$ is chosen in a way that the $\nu=2$ minimum is finite and should be sensitive to any increase in temperature. The data in panel \textbf{c} shows $\rho$ increases by a factor of approximately three when increasing the current to 5~nA. The data taken at zero field (blue) reveal a notably similar character. 

\begin{figure}[h]
	\includegraphics[width=56mm]{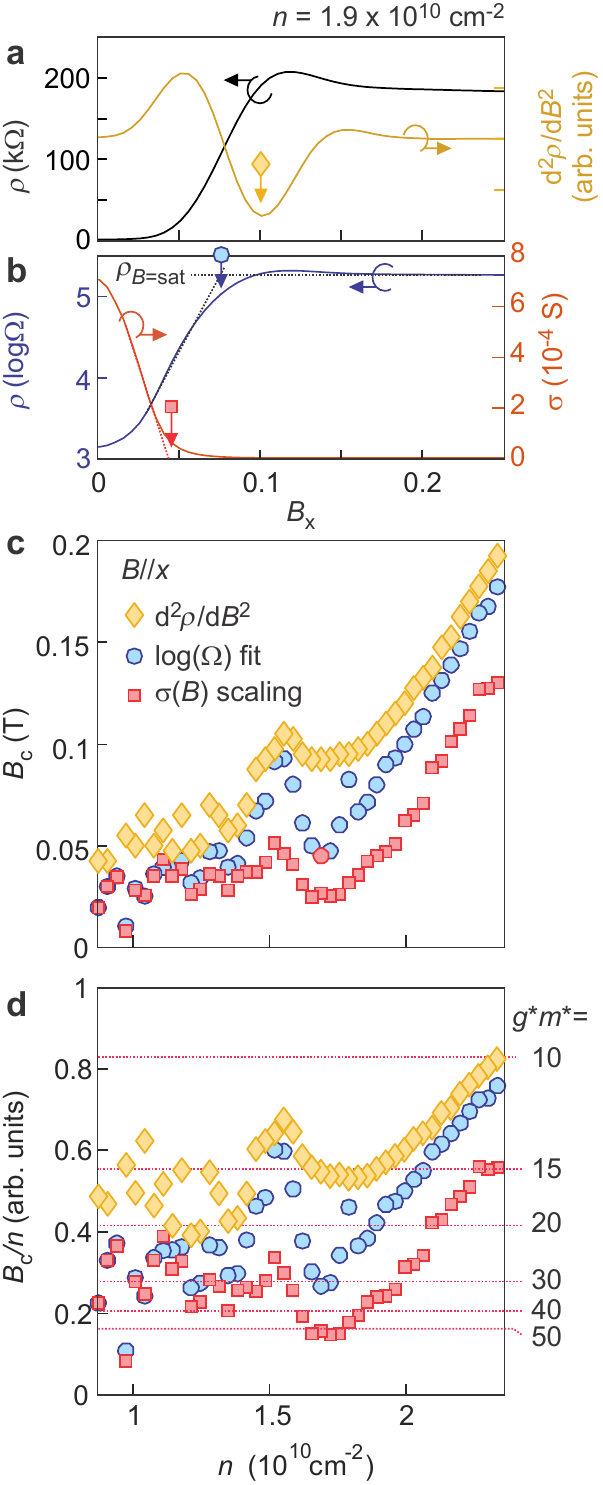}
	\caption{\textbf{Comparison of $\Bc$ from multiple approaches.} Here, $n=1.9$\density. \textbf{a} $\rho$ and $d^2\rho/dB^2$ as a function of $\Bx$. The orange diamond highlights the minimum in the second derivative used to quantify $\Bc$. \textbf{b} $\rho$ plotted on a log scale and $\sigma(B)$ as a function of $\Bx$. The blue circle and red square highlight $\Bc$ for these respective traces. \textbf{c} $\Bc$ and \textbf{d} ratio of $\Bc/n$ as a function of $n$ for these three approaches. The corresponding $g^*m^*$ is indicated to the right hand side of panel \textbf{d}. $T\approx$10~mK for all data.}
	\label{S_chiestimate}
\end{figure}

At this charge density ($n=1.8$\density), $\rho$ reaches 10~k$\Omega$ when we heat the mixing chamber to be $T_\mathrm{MC}\approx80$~mK. A heating power approaching 300~$\Mu$W applied to the mixing chamber plate is required in order to achieve that temperature at the thermometer. Panel \textbf{d} takes $\rho$ and plots it as a function of dissipated power across voltage probes ($P=IV$). It is difficult to predict the exact relationship between $\rho$ and power owing to the complicated temperature dependence of the 2DES and thermal coupling of the device to cryogenics surfaces. We remark that the power dissipated inside the samples is 10 orders of magnitude smaller than the equivalent power required at the mixing chamber plate. While our transport measurements preclude detailed conclusions about the texture of carriers inside the device, we speculate that conventional Joule heating of a homogeneous liquid could not be responsible for this increase in $\rho$. 

\section{Methods for determining $\Bc$}

The uncertainty involved in the quantification of $\Bc$~at very low field values results in deviations in the calculated value of $\chi$. This uncertainty has quantitative impacts on the results discussed in Fig.~\ref{Fig3}, but does not change the overall physical picture discussed in this manuscript. For the sake of completeness, here we discuss identification of $\Bc$ using three different methods. The second derivative of $\rho$ with respect to $B$ exhibits a local minimum in the vicinity of $\Bc$, which can be used to identify its value. This approach may be systematically applied to analyze the data and obtain $\Bc$($n$). An example of the raw data and second derivative is presented in Fig.~\ref{S_chiestimate}\textbf{a}. It is visually apparent that the local minimum in $d^2r/dB^2$ occurs at a value of $B_x$ that is close to where the positive magnetoresistance saturates.  A second method for estimating $\Bc$ is based on fitting a linear slope to $\log \rho(B)$ in the positive magnetoresistance regime. $\Bc$ is defined as the point at which the extrapolated value of this slope is equal to the saturated resistance at high $B$ ($\rho_\mathrm{B=sat}$). An example is displayed in Fig.~\ref{S_chiestimate}\textbf{b} as the blue trace. Finally, the framework in Ref.~[\onlinecite{vitkalov:2001}] provides a third method for extracting $\Bc$ using the magnetoconductivity data. In this method, $B_c$ is identified with the crossover between a linear fit of $\sigma(B)$ and the value $\sigma_\infty$ at $\Bx \rightarrow \infty$. This third method is demonstrated in Fig.~\ref{S_chiestimate}\textbf{b} via the red trace. 

The results of these three quantification approaches are displayed in Fig.~\ref{S_chiestimate}\textbf{c} and \textbf{d}. While some systematic discrepancy is evident between the three methods, they each produce the same qualitative features, including: 1) a steady decrease (increase) of $\Bc$ ($g^*m^*$) as $n$ is reduced from above  $\nc$, 2) an alteration of this trend at approximately $n=\nc$, and 3) an apparent saturation of $\Bc/n$ as when $n\ll\nc$.

\begin{figure}
	\includegraphics[width=84mm]{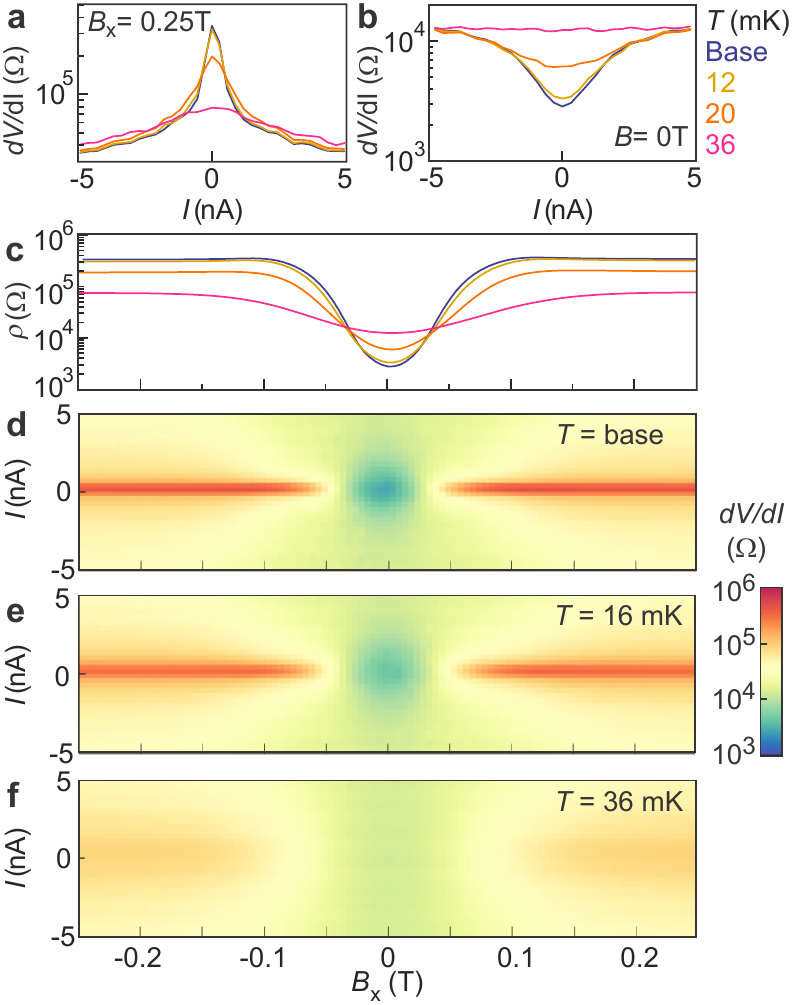}
	\caption{\textbf{Examination of excess conductivity.} Here, $n$=1.82\density. Temperature dependence of $dV/dI$ as a function of $I$ at \textbf{a} $\Bx$=0.25~T and \textbf{b} $\Bx$=0~T. \textbf{c} $\rho$ as a function of $\Bx$. The measurement temperature is indicated outside the right hand side of panel \textbf{b}. Panels \textbf{d-f} map $dV/dI$ in the ($\Bx$,$I$)-plane at three distinct temperatures.}
	\label{S_EC}
\end{figure}

\begin{figure}
	\includegraphics[width=82mm]{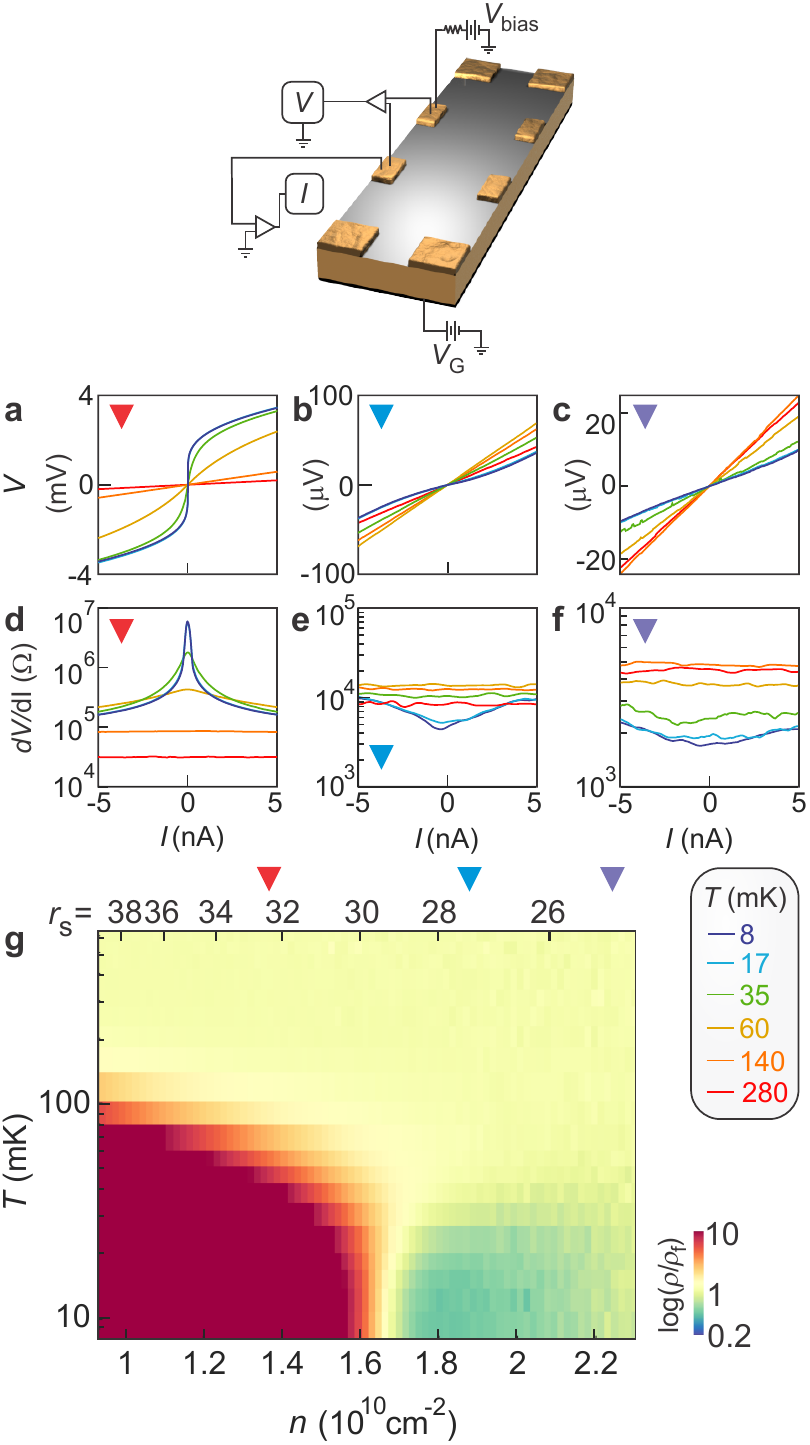}
	\caption{\textbf{Charge transport characteristics in a two-point measurement geometry.} \textbf{a-c} $I$-$V$ sweeps at various $T$ when $n$=1.38, 1.91, and 2.26\density. \textbf{d-f} First differential ($dV/dI$) of the $I$-$V$ characteristics for the same set of temperatures. \textbf{g} Non-linearity, defined as $\rho/\rho_\mathrm{f}$ in the current-voltage characteristics of the device as a function of $n$ and $T$. Colored triangles highlight the three regimes discussed in the text and their location on panel \textbf{g}.}
	\label{S_twopoint}
\end{figure}

\section{Temperature and field dependence of excess conductance}

The excess conductance identified in Fig.~\ref{Fig2} bears some resemblance to other anomalous metal phases discussed in the literature.\cite{kapitulnik:2019} In such systems, the longitudinal resistivity falls at low temperature as if approaching a superconducting ground state, only to saturate at a finite value. This saturation is accompanied by strong positive magnetoresistance. Our results, along with others present in other studies,\cite{kravchenko:1996,yoon:1999} identify this excess conductance occurring immediately prior to the MIT at zero field. Figure~\ref{S_EC} presents an extended data set taken at $n=1.82$\density~in the device reported in this work. Panels \textbf{a} and \textbf{b} present $dV/dI$ at various temperatures as a function of $I$ at $B=0.25$ and 0~T, respectively. A transition from a zero-electric-field insulating state to the excess conductance phase is evident between these two conditions. The former occurs when the 2DES is fully spin polarized, as discussed above. This is clear based on an examination of panel \textbf{c}, where $\rho$ is plotted as a function of $\Bx$, and is seen to saturate above $|B_\mathrm{x}|\approx 0.07$~T. Panels \textbf{d-f} map $dV/dI$ in the ($\Bx$,$I$)--plane at three distinct temperatures. The excess conductance phase is observed at low $|B|$ and $|I|$ as a blue region at low temperature. The insulating phase emerges above $\Bc$~and is restricted to low $|I|$. By $T=36$~mK both of these features are largely washed out. 

\section{Two-point current-voltage characteristics}

Data presented throughout the main text and previous supplementary information sections have all been gathered in a four-point measurement geometry, as shown in Fig.\ref{Fig1}\textbf{a}. Figure \ref{S_twopoint} presents data from the same experimental procedure performed to gather the data of Fig.~\ref{Fig2}, only taken in a two-point configuration. Here, current is supplied to the device through the sample contacts used to measure the voltage drop, as shown in the upper schematic of Fig.~\ref{S_twopoint}. The voltage preamplifier is connected after the load resistor ($R=10$~M$\Omega$) at the break-out box. The signal therefore contains a contribution from the measurement wires (approximately 150$\Omega$ each direction) and contact resistance of the device. The data in Fig.~\ref{S_twopoint} qualitatively reproduce the result presented in Fig.\ref{Fig4} of the main text. The data confirms that the contacts are ohmic for all $n$ when $T>100$\,mK.


\begin{thebibliography}{51}
	\expandafter\ifx\csname natexlab\endcsname\relax\def\natexlab#1{#1}\fi
	\expandafter\ifx\csname bibnamefont\endcsname\relax
	\def\bibnamefont#1{#1}\fi
	\expandafter\ifx\csname bibfnamefont\endcsname\relax
	\def\bibfnamefont#1{#1}\fi
	\expandafter\ifx\csname citenamefont\endcsname\relax
	\def\citenamefont#1{#1}\fi
	\expandafter\ifx\csname url\endcsname\relax
	\def\url#1{\texttt{#1}}\fi
	\expandafter\ifx\csname urlprefix\endcsname\relax\def\urlprefix{URL }\fi
	\providecommand{\bibinfo}[2]{#2}
	\providecommand{\eprint}[2][]{\url{#2}}
	
	\bibitem[{\citenamefont{Tanatar and Ceperley}(1989)}]{tanatar:1989}
	\bibinfo{author}{\bibfnamefont{B.}~\bibnamefont{Tanatar}} \bibnamefont{and}
	\bibinfo{author}{\bibfnamefont{D.~M.} \bibnamefont{Ceperley}},
	\bibinfo{journal}{Phys. Rev. B} \textbf{\bibinfo{volume}{39}},
	\bibinfo{pages}{5005} (\bibinfo{year}{1989}),
	\urlprefix\url{https://link.aps.org/doi/10.1103/PhysRevB.39.5005}.
	
	\bibitem[{\citenamefont{Rapisarda and Senatore}(1996)}]{rapisarda:1996}
	\bibinfo{author}{\bibfnamefont{F.}~\bibnamefont{Rapisarda}} \bibnamefont{and}
	\bibinfo{author}{\bibfnamefont{G.}~\bibnamefont{Senatore}},
	\bibinfo{journal}{Australian journal of physics}
	\textbf{\bibinfo{volume}{49}}, \bibinfo{pages}{161} (\bibinfo{year}{1996}).
	
	\bibitem[{\citenamefont{Phillips et~al.}(1998)\citenamefont{Phillips, Wan,
			Martin, Knysh, and Dalidovich}}]{phillips:1998}
	\bibinfo{author}{\bibfnamefont{P.}~\bibnamefont{Phillips}},
	\bibinfo{author}{\bibfnamefont{Y.}~\bibnamefont{Wan}},
	\bibinfo{author}{\bibfnamefont{I.}~\bibnamefont{Martin}},
	\bibinfo{author}{\bibfnamefont{S.}~\bibnamefont{Knysh}}, \bibnamefont{and}
	\bibinfo{author}{\bibfnamefont{D.}~\bibnamefont{Dalidovich}},
	\bibinfo{journal}{Nature} \textbf{\bibinfo{volume}{395}},
	\bibinfo{pages}{253} (\bibinfo{year}{1998}).
	
	\bibitem[{\citenamefont{Chamon et~al.}(2001)\citenamefont{Chamon, Mucciolo, and
			Castro~Neto}}]{chamon:2001}
	\bibinfo{author}{\bibfnamefont{C.}~\bibnamefont{Chamon}},
	\bibinfo{author}{\bibfnamefont{E.~R.} \bibnamefont{Mucciolo}},
	\bibnamefont{and} \bibinfo{author}{\bibfnamefont{A.~H.}
		\bibnamefont{Castro~Neto}}, \bibinfo{journal}{Phys. Rev. B}
	\textbf{\bibinfo{volume}{64}}, \bibinfo{pages}{245115}
	(\bibinfo{year}{2001}),
	\urlprefix\url{https://link.aps.org/doi/10.1103/PhysRevB.64.245115}.
	
	\bibitem[{\citenamefont{Attaccalite et~al.}(2002)\citenamefont{Attaccalite,
			Moroni, Gori-Giorgi, and Bachelet}}]{attaccalite:2002}
	\bibinfo{author}{\bibfnamefont{C.}~\bibnamefont{Attaccalite}},
	\bibinfo{author}{\bibfnamefont{S.}~\bibnamefont{Moroni}},
	\bibinfo{author}{\bibfnamefont{P.}~\bibnamefont{Gori-Giorgi}},
	\bibnamefont{and} \bibinfo{author}{\bibfnamefont{G.~B.}
		\bibnamefont{Bachelet}}, \bibinfo{journal}{Phys. Rev. Lett.}
	\textbf{\bibinfo{volume}{88}}, \bibinfo{pages}{256601}
	(\bibinfo{year}{2002}),
	\urlprefix\url{https://link.aps.org/doi/10.1103/PhysRevLett.88.256601}.
	
	\bibitem[{\citenamefont{Spivak and Kivelson}(2004)}]{spivak:2004}
	\bibinfo{author}{\bibfnamefont{B.}~\bibnamefont{Spivak}} \bibnamefont{and}
	\bibinfo{author}{\bibfnamefont{S.~A.} \bibnamefont{Kivelson}},
	\bibinfo{journal}{Phys. Rev. B} \textbf{\bibinfo{volume}{70}},
	\bibinfo{pages}{155114} (\bibinfo{year}{2004}),
	\urlprefix\url{https://link.aps.org/doi/10.1103/PhysRevB.70.155114}.
	
	\bibitem[{\citenamefont{Drummond and Needs}(2009)}]{drummond:2009}
	\bibinfo{author}{\bibfnamefont{N.~D.} \bibnamefont{Drummond}} \bibnamefont{and}
	\bibinfo{author}{\bibfnamefont{R.~J.} \bibnamefont{Needs}},
	\bibinfo{journal}{Phys. Rev. Lett.} \textbf{\bibinfo{volume}{102}},
	\bibinfo{pages}{126402} (\bibinfo{year}{2009}),
	\urlprefix\url{https://link.aps.org/doi/10.1103/PhysRevLett.102.126402}.
	
	\bibitem[{\citenamefont{Spivak et~al.}(2010)\citenamefont{Spivak, Kravchenko,
			Kivelson, and Gao}}]{spivak:2010}
	\bibinfo{author}{\bibfnamefont{B.}~\bibnamefont{Spivak}},
	\bibinfo{author}{\bibfnamefont{S.~V.} \bibnamefont{Kravchenko}},
	\bibinfo{author}{\bibfnamefont{S.~A.} \bibnamefont{Kivelson}},
	\bibnamefont{and} \bibinfo{author}{\bibfnamefont{X.~P.~A.}
		\bibnamefont{Gao}}, \bibinfo{journal}{Rev. Mod. Phys.}
	\textbf{\bibinfo{volume}{82}}, \bibinfo{pages}{1743} (\bibinfo{year}{2010}),
	\urlprefix\url{https://link.aps.org/doi/10.1103/RevModPhys.82.1743}.
	
	\bibitem[{\citenamefont{Abrahams et~al.}(2001)\citenamefont{Abrahams,
			Kravchenko, and Sarachik}}]{abrahams:2001}
	\bibinfo{author}{\bibfnamefont{E.}~\bibnamefont{Abrahams}},
	\bibinfo{author}{\bibfnamefont{S.~V.} \bibnamefont{Kravchenko}},
	\bibnamefont{and} \bibinfo{author}{\bibfnamefont{M.~P.}
		\bibnamefont{Sarachik}}, \bibinfo{journal}{Rev. Mod. Phys.}
	\textbf{\bibinfo{volume}{73}}, \bibinfo{pages}{251} (\bibinfo{year}{2001}),
	\urlprefix\url{https://link.aps.org/doi/10.1103/RevModPhys.73.251}.
	
	\bibitem[{\citenamefont{Kravchenko and Sarachik}(2003)}]{Kravchenko:2003}
	\bibinfo{author}{\bibfnamefont{S.~V.} \bibnamefont{Kravchenko}}
	\bibnamefont{and} \bibinfo{author}{\bibfnamefont{M.~P.}
		\bibnamefont{Sarachik}}, \bibinfo{journal}{Reports on Progress in Physics}
	\textbf{\bibinfo{volume}{67}}, \bibinfo{pages}{1} (\bibinfo{year}{2003}),
	\urlprefix\url{https://doi.org/10.1088/0034-4885/67/1/r01}.
	
	\bibitem[{\citenamefont{Shashkin and Kravchenko}(2019)}]{shashkin:2019}
	\bibinfo{author}{\bibfnamefont{A.~A.} \bibnamefont{Shashkin}} \bibnamefont{and}
	\bibinfo{author}{\bibfnamefont{S.~V.} \bibnamefont{Kravchenko}},
	\bibinfo{journal}{Applied Sciences} \textbf{\bibinfo{volume}{9}}
	(\bibinfo{year}{2019}), ISSN \bibinfo{issn}{2076-3417},
	\urlprefix\url{https://www.mdpi.com/2076-3417/9/6/1169}.
	
	\bibitem[{\citenamefont{Dolgopolov}(2019)}]{dolgopolov:2019}
	\bibinfo{author}{\bibfnamefont{V.~T.} \bibnamefont{Dolgopolov}},
	\bibinfo{journal}{Physics-Uspekhi} \textbf{\bibinfo{volume}{62}},
	\bibinfo{pages}{633} (\bibinfo{year}{2019}),
	\urlprefix\url{https://doi.org/10.3367/ufne.2018.10.038449}.
	
	\bibitem[{\citenamefont{Yoon et~al.}(1999)\citenamefont{Yoon, Li, Shahar, Tsui,
			and Shayegan}}]{yoon:1999}
	\bibinfo{author}{\bibfnamefont{J.}~\bibnamefont{Yoon}},
	\bibinfo{author}{\bibfnamefont{C.~C.} \bibnamefont{Li}},
	\bibinfo{author}{\bibfnamefont{D.}~\bibnamefont{Shahar}},
	\bibinfo{author}{\bibfnamefont{D.~C.} \bibnamefont{Tsui}}, \bibnamefont{and}
	\bibinfo{author}{\bibfnamefont{M.}~\bibnamefont{Shayegan}},
	\bibinfo{journal}{Phys. Rev. Lett.} \textbf{\bibinfo{volume}{82}},
	\bibinfo{pages}{1744} (\bibinfo{year}{1999}),
	\urlprefix\url{https://link.aps.org/doi/10.1103/PhysRevLett.82.1744}.
	
	\bibitem[{\citenamefont{Knighton et~al.}(2018)\citenamefont{Knighton, Wu,
			Huang, Serafin, Xia, Pfeiffer, and West}}]{knighton:2018}
	\bibinfo{author}{\bibfnamefont{T.}~\bibnamefont{Knighton}},
	\bibinfo{author}{\bibfnamefont{Z.}~\bibnamefont{Wu}},
	\bibinfo{author}{\bibfnamefont{J.}~\bibnamefont{Huang}},
	\bibinfo{author}{\bibfnamefont{A.}~\bibnamefont{Serafin}},
	\bibinfo{author}{\bibfnamefont{J.~S.} \bibnamefont{Xia}},
	\bibinfo{author}{\bibfnamefont{L.~N.} \bibnamefont{Pfeiffer}},
	\bibnamefont{and} \bibinfo{author}{\bibfnamefont{K.~W.} \bibnamefont{West}},
	\bibinfo{journal}{Phys. Rev. B} \textbf{\bibinfo{volume}{97}},
	\bibinfo{pages}{085135} (\bibinfo{year}{2018}),
	\urlprefix\url{https://link.aps.org/doi/10.1103/PhysRevB.97.085135}.
	
	\bibitem[{\citenamefont{Hossain et~al.}(2020)\citenamefont{Hossain, Ma,
			Rosales, Chung, Pfeiffer, West, Baldwin, and Shayegan}}]{hossain:2020}
	\bibinfo{author}{\bibfnamefont{M.~S.} \bibnamefont{Hossain}},
	\bibinfo{author}{\bibfnamefont{M.~K.} \bibnamefont{Ma}},
	\bibinfo{author}{\bibfnamefont{K.~A.~V.} \bibnamefont{Rosales}},
	\bibinfo{author}{\bibfnamefont{Y.~J.} \bibnamefont{Chung}},
	\bibinfo{author}{\bibfnamefont{L.~N.} \bibnamefont{Pfeiffer}},
	\bibinfo{author}{\bibfnamefont{K.~W.} \bibnamefont{West}},
	\bibinfo{author}{\bibfnamefont{K.~W.} \bibnamefont{Baldwin}},
	\bibnamefont{and} \bibinfo{author}{\bibfnamefont{M.}~\bibnamefont{Shayegan}},
	\bibinfo{journal}{Proceedings of the National Academy of Sciences}
	\textbf{\bibinfo{volume}{117}}, \bibinfo{pages}{32244}
	(\bibinfo{year}{2020}), ISSN \bibinfo{issn}{0027-8424},
	\eprint{https://www.pnas.org/content/117/51/32244.full.pdf},
	\urlprefix\url{https://www.pnas.org/content/117/51/32244}.
	
	\bibitem[{\citenamefont{Falson et~al.}(2015)\citenamefont{Falson, Maryenko,
			Friess, Zhang, Kozuka, Tsukazaki, Smet, and Kawasaki}}]{falson:2015a}
	\bibinfo{author}{\bibfnamefont{J.}~\bibnamefont{Falson}},
	\bibinfo{author}{\bibfnamefont{D.}~\bibnamefont{Maryenko}},
	\bibinfo{author}{\bibfnamefont{B.}~\bibnamefont{Friess}},
	\bibinfo{author}{\bibfnamefont{D.}~\bibnamefont{Zhang}},
	\bibinfo{author}{\bibfnamefont{Y.}~\bibnamefont{Kozuka}},
	\bibinfo{author}{\bibfnamefont{A.}~\bibnamefont{Tsukazaki}},
	\bibinfo{author}{\bibfnamefont{J.~H.} \bibnamefont{Smet}}, \bibnamefont{and}
	\bibinfo{author}{\bibfnamefont{M.}~\bibnamefont{Kawasaki}},
	\bibinfo{journal}{Nature Physics} \textbf{\bibinfo{volume}{11}},
	\bibinfo{pages}{347} (\bibinfo{year}{2015}).
	
	\bibitem[{\citenamefont{Falson et~al.}(2018)\citenamefont{Falson, Tabrea,
			Zhang, Sodemann, Kozuka, Tsukazaki, Kawasaki, von Klitzing, and
			Smet}}]{falson:2018b}
	\bibinfo{author}{\bibfnamefont{J.}~\bibnamefont{Falson}},
	\bibinfo{author}{\bibfnamefont{D.}~\bibnamefont{Tabrea}},
	\bibinfo{author}{\bibfnamefont{D.}~\bibnamefont{Zhang}},
	\bibinfo{author}{\bibfnamefont{I.}~\bibnamefont{Sodemann}},
	\bibinfo{author}{\bibfnamefont{Y.}~\bibnamefont{Kozuka}},
	\bibinfo{author}{\bibfnamefont{A.}~\bibnamefont{Tsukazaki}},
	\bibinfo{author}{\bibfnamefont{M.}~\bibnamefont{Kawasaki}},
	\bibinfo{author}{\bibfnamefont{K.}~\bibnamefont{von Klitzing}},
	\bibnamefont{and} \bibinfo{author}{\bibfnamefont{J.~H.} \bibnamefont{Smet}},
	\bibinfo{journal}{Science Advances} \textbf{\bibinfo{volume}{4}},
	\bibinfo{pages}{eaat8742} (\bibinfo{year}{2018}).
	
	\bibitem[{\citenamefont{Kozuka et~al.}(2013)\citenamefont{Kozuka, Teraoka,
			Falson, Oiwa, Tsukazaki, Tarucha, and Kawasaki}}]{kozuka:2013}
	\bibinfo{author}{\bibfnamefont{Y.}~\bibnamefont{Kozuka}},
	\bibinfo{author}{\bibfnamefont{S.}~\bibnamefont{Teraoka}},
	\bibinfo{author}{\bibfnamefont{J.}~\bibnamefont{Falson}},
	\bibinfo{author}{\bibfnamefont{A.}~\bibnamefont{Oiwa}},
	\bibinfo{author}{\bibfnamefont{A.}~\bibnamefont{Tsukazaki}},
	\bibinfo{author}{\bibfnamefont{S.}~\bibnamefont{Tarucha}}, \bibnamefont{and}
	\bibinfo{author}{\bibfnamefont{M.}~\bibnamefont{Kawasaki}},
	\bibinfo{journal}{Phys. Rev. B} \textbf{\bibinfo{volume}{87}},
	\bibinfo{pages}{205411} (\bibinfo{year}{2013}),
	\urlprefix\url{https://link.aps.org/doi/10.1103/PhysRevB.87.205411}.
	
	\bibitem[{\citenamefont{Shklovskii and
			Efros}(1984)}]{shklovskii_electronic_1984}
	\bibinfo{author}{\bibfnamefont{B.~I.} \bibnamefont{Shklovskii}}
	\bibnamefont{and} \bibinfo{author}{\bibfnamefont{A.~L.} \bibnamefont{Efros}},
	\emph{\bibinfo{title}{Electronic {Properties} of {Doped} {Semiconductors}}}
	(\bibinfo{publisher}{Springer-Verlag}, \bibinfo{address}{New York},
	\bibinfo{year}{1984}).
	
	\bibitem[{\citenamefont{Shklovskii}(2004)}]{shklovskii:2004}
	\bibinfo{author}{\bibfnamefont{B.~I.} \bibnamefont{Shklovskii}},
	\bibinfo{journal}{physica status solidi (c)} \textbf{\bibinfo{volume}{1}},
	\bibinfo{pages}{46} (\bibinfo{year}{2004}),
	\eprint{https://onlinelibrary.wiley.com/doi/pdf/10.1002/pssc.200303642},
	\urlprefix\url{https://onlinelibrary.wiley.com/doi/abs/10.1002/pssc.200303642}.
	
	\bibitem[{\citenamefont{Zala et~al.}(2001{\natexlab{a}})\citenamefont{Zala,
			Narozhny, and Aleiner}}]{Zala2001rapid}
	\bibinfo{author}{\bibfnamefont{G.}~\bibnamefont{Zala}},
	\bibinfo{author}{\bibfnamefont{B.~N.} \bibnamefont{Narozhny}},
	\bibnamefont{and} \bibinfo{author}{\bibfnamefont{I.~L.}
		\bibnamefont{Aleiner}}, \bibinfo{journal}{Phys. Rev. B}
	\textbf{\bibinfo{volume}{65}}, \bibinfo{pages}{020201}
	(\bibinfo{year}{2001}{\natexlab{a}}),
	\urlprefix\url{https://link.aps.org/doi/10.1103/PhysRevB.65.020201}.
	
	\bibitem[{\citenamefont{Zala et~al.}(2001{\natexlab{b}})\citenamefont{Zala,
			Narozhny, and Aleiner}}]{Zala2001interaction}
	\bibinfo{author}{\bibfnamefont{G.}~\bibnamefont{Zala}},
	\bibinfo{author}{\bibfnamefont{B.~N.} \bibnamefont{Narozhny}},
	\bibnamefont{and} \bibinfo{author}{\bibfnamefont{I.~L.}
		\bibnamefont{Aleiner}}, \bibinfo{journal}{Phys. Rev. B}
	\textbf{\bibinfo{volume}{64}}, \bibinfo{pages}{214204}
	(\bibinfo{year}{2001}{\natexlab{b}}),
	\urlprefix\url{https://link.aps.org/doi/10.1103/PhysRevB.64.214204}.
	
	\bibitem[{\citenamefont{Kravchenko et~al.}(1996)\citenamefont{Kravchenko,
			Simonian, Sarachik, Mason, and Furneaux}}]{kravchenko:1996}
	\bibinfo{author}{\bibfnamefont{S.~V.} \bibnamefont{Kravchenko}},
	\bibinfo{author}{\bibfnamefont{D.}~\bibnamefont{Simonian}},
	\bibinfo{author}{\bibfnamefont{M.~P.} \bibnamefont{Sarachik}},
	\bibinfo{author}{\bibfnamefont{W.}~\bibnamefont{Mason}}, \bibnamefont{and}
	\bibinfo{author}{\bibfnamefont{J.~E.} \bibnamefont{Furneaux}},
	\bibinfo{journal}{Phys. Rev. Lett.} \textbf{\bibinfo{volume}{77}},
	\bibinfo{pages}{4938} (\bibinfo{year}{1996}),
	\urlprefix\url{https://link.aps.org/doi/10.1103/PhysRevLett.77.4938}.
	
	\bibitem[{\citenamefont{Matveev et~al.}(1995)\citenamefont{Matveev, Glazman,
			Clarke, Ephron, and Beasley}}]{Matveev1995}
	\bibinfo{author}{\bibfnamefont{K.~A.} \bibnamefont{Matveev}},
	\bibinfo{author}{\bibfnamefont{L.~I.} \bibnamefont{Glazman}},
	\bibinfo{author}{\bibfnamefont{P.}~\bibnamefont{Clarke}},
	\bibinfo{author}{\bibfnamefont{D.}~\bibnamefont{Ephron}}, \bibnamefont{and}
	\bibinfo{author}{\bibfnamefont{M.~R.} \bibnamefont{Beasley}},
	\bibinfo{journal}{Phys. Rev. B} \textbf{\bibinfo{volume}{52}},
	\bibinfo{pages}{5289} (\bibinfo{year}{1995}),
	\urlprefix\url{https://link.aps.org/doi/10.1103/PhysRevB.52.5289}.
	
	\bibitem[{\citenamefont{Chakravarty et~al.}(1999)\citenamefont{Chakravarty,
			Kivelson, Nayak, and Voelker}}]{chakravarty:1999}
	\bibinfo{author}{\bibfnamefont{S.}~\bibnamefont{Chakravarty}},
	\bibinfo{author}{\bibfnamefont{S.}~\bibnamefont{Kivelson}},
	\bibinfo{author}{\bibfnamefont{C.}~\bibnamefont{Nayak}}, \bibnamefont{and}
	\bibinfo{author}{\bibfnamefont{K.}~\bibnamefont{Voelker}},
	\bibinfo{journal}{Philosophical Magazine B} \textbf{\bibinfo{volume}{79}},
	\bibinfo{pages}{859} (\bibinfo{year}{1999}),
	\eprint{https://doi.org/10.1080/13642819908214845},
	\urlprefix\url{https://doi.org/10.1080/13642819908214845}.
	
	\bibitem[{\citenamefont{Bernu et~al.}(2001)\citenamefont{Bernu, C\^andido, and
			Ceperley}}]{bernu:2001}
	\bibinfo{author}{\bibfnamefont{B.}~\bibnamefont{Bernu}},
	\bibinfo{author}{\bibfnamefont{L.}~\bibnamefont{C\^andido}},
	\bibnamefont{and} \bibinfo{author}{\bibfnamefont{D.~M.}
		\bibnamefont{Ceperley}}, \bibinfo{journal}{Phys. Rev. Lett.}
	\textbf{\bibinfo{volume}{86}}, \bibinfo{pages}{870} (\bibinfo{year}{2001}),
	\urlprefix\url{https://link.aps.org/doi/10.1103/PhysRevLett.86.870}.
	
	\bibitem[{\citenamefont{Bernu et~al.}(2011)\citenamefont{Bernu, Delyon,
			Holzmann, and Baguet}}]{bernu:2011}
	\bibinfo{author}{\bibfnamefont{B.}~\bibnamefont{Bernu}},
	\bibinfo{author}{\bibfnamefont{F.}~\bibnamefont{Delyon}},
	\bibinfo{author}{\bibfnamefont{M.}~\bibnamefont{Holzmann}}, \bibnamefont{and}
	\bibinfo{author}{\bibfnamefont{L.}~\bibnamefont{Baguet}},
	\bibinfo{journal}{Phys. Rev. B} \textbf{\bibinfo{volume}{84}},
	\bibinfo{pages}{115115} (\bibinfo{year}{2011}),
	\urlprefix\url{https://link.aps.org/doi/10.1103/PhysRevB.84.115115}.
	
	\bibitem[{\citenamefont{Bernu et~al.}(2017)\citenamefont{Bernu, Delyon, Baguet,
			and Holzmann}}]{bernu:2017}
	\bibinfo{author}{\bibfnamefont{B.}~\bibnamefont{Bernu}},
	\bibinfo{author}{\bibfnamefont{F.}~\bibnamefont{Delyon}},
	\bibinfo{author}{\bibfnamefont{L.}~\bibnamefont{Baguet}}, \bibnamefont{and}
	\bibinfo{author}{\bibfnamefont{M.}~\bibnamefont{Holzmann}},
	\bibinfo{journal}{Contributions to Plasma Physics}
	\textbf{\bibinfo{volume}{57}}, \bibinfo{pages}{524} (\bibinfo{year}{2017}),
	\eprint{https://onlinelibrary.wiley.com/doi/pdf/10.1002/ctpp.201700139},
	\urlprefix\url{https://onlinelibrary.wiley.com/doi/abs/10.1002/ctpp.201700139}.
	
	\bibitem[{\citenamefont{Jamei et~al.}(2005)\citenamefont{Jamei, Kivelson, and
			Spivak}}]{jamel:2005}
	\bibinfo{author}{\bibfnamefont{R.}~\bibnamefont{Jamei}},
	\bibinfo{author}{\bibfnamefont{S.}~\bibnamefont{Kivelson}}, \bibnamefont{and}
	\bibinfo{author}{\bibfnamefont{B.}~\bibnamefont{Spivak}},
	\bibinfo{journal}{Phys. Rev. Lett.} \textbf{\bibinfo{volume}{94}},
	\bibinfo{pages}{056805} (\bibinfo{year}{2005}),
	\urlprefix\url{https://link.aps.org/doi/10.1103/PhysRevLett.94.056805}.
	
	\bibitem[{\citenamefont{Spivak and Kivelson}(2006)}]{spivak:2006}
	\bibinfo{author}{\bibfnamefont{B.}~\bibnamefont{Spivak}} \bibnamefont{and}
	\bibinfo{author}{\bibfnamefont{S.~A.} \bibnamefont{Kivelson}},
	\bibinfo{journal}{Annals of Physics} \textbf{\bibinfo{volume}{321}},
	\bibinfo{pages}{2071 } (\bibinfo{year}{2006}), ISSN
	\bibinfo{issn}{0003-4916},
	\urlprefix\url{http://www.sciencedirect.com/science/article/pii/S0003491605002654}.
	
	\bibitem[{\citenamefont{Li et~al.}(2019)\citenamefont{Li, Zhang, Ghaemi, and
			Sarachik}}]{li:2019}
	\bibinfo{author}{\bibfnamefont{S.}~\bibnamefont{Li}},
	\bibinfo{author}{\bibfnamefont{Q.}~\bibnamefont{Zhang}},
	\bibinfo{author}{\bibfnamefont{P.}~\bibnamefont{Ghaemi}}, \bibnamefont{and}
	\bibinfo{author}{\bibfnamefont{M.~P.} \bibnamefont{Sarachik}},
	\bibinfo{journal}{Phys. Rev. B} \textbf{\bibinfo{volume}{99}},
	\bibinfo{pages}{155302} (\bibinfo{year}{2019}),
	\urlprefix\url{https://link.aps.org/doi/10.1103/PhysRevB.99.155302}.
	
	\bibitem[{\citenamefont{{J. Falson and Y. Kozuka and M. Uchida and J. H Smet
				and T. Arima and A. Tsukazaki and M. Kawasaki}}(2016)}]{falson:2016}
	\bibinfo{author}{\bibnamefont{{J. Falson and Y. Kozuka and M. Uchida and J. H
				Smet and T. Arima and A. Tsukazaki and M. Kawasaki}}},
	\bibinfo{journal}{Scientific reports} \textbf{\bibinfo{volume}{6}},
	\bibinfo{pages}{26598} (\bibinfo{year}{2016}).
	
	\bibitem[{\citenamefont{Falson and Kawasaki}(2018)}]{falson:2018a}
	\bibinfo{author}{\bibfnamefont{J.}~\bibnamefont{Falson}} \bibnamefont{and}
	\bibinfo{author}{\bibfnamefont{M.}~\bibnamefont{Kawasaki}},
	\bibinfo{journal}{Reports on Progress in Physics}
	\textbf{\bibinfo{volume}{81}}, \bibinfo{pages}{056501}
	(\bibinfo{year}{2018}).
	
	\bibitem[{\citenamefont{Pan et~al.}(1999)\citenamefont{Pan, Xia, Shvarts,
			Adams, Stormer, Tsui, Pfeiffer, Baldwin, and West}}]{pan:1999}
	\bibinfo{author}{\bibfnamefont{W.}~\bibnamefont{Pan}},
	\bibinfo{author}{\bibfnamefont{J.-S.} \bibnamefont{Xia}},
	\bibinfo{author}{\bibfnamefont{V.}~\bibnamefont{Shvarts}},
	\bibinfo{author}{\bibfnamefont{D.~E.} \bibnamefont{Adams}},
	\bibinfo{author}{\bibfnamefont{H.~L.} \bibnamefont{Stormer}},
	\bibinfo{author}{\bibfnamefont{D.~C.} \bibnamefont{Tsui}},
	\bibinfo{author}{\bibfnamefont{L.~N.} \bibnamefont{Pfeiffer}},
	\bibinfo{author}{\bibfnamefont{K.~W.} \bibnamefont{Baldwin}},
	\bibnamefont{and} \bibinfo{author}{\bibfnamefont{K.~W.} \bibnamefont{West}},
	\bibinfo{journal}{Phys. Rev. Lett.} \textbf{\bibinfo{volume}{83}},
	\bibinfo{pages}{3530} (\bibinfo{year}{1999}),
	\urlprefix\url{https://link.aps.org/doi/10.1103/PhysRevLett.83.3530}.
	
	\bibitem[{\citenamefont{Solovyev et~al.}(2015)\citenamefont{Solovyev, Van'kov,
			Kukushkin, Falson, Zhang, Maryenko, Kozuka, Tsukazaki, Smet, and
			Kawasaki}}]{solovyev:2015}
	\bibinfo{author}{\bibfnamefont{V.~V.} \bibnamefont{Solovyev}},
	\bibinfo{author}{\bibfnamefont{A.~B.} \bibnamefont{Van'kov}},
	\bibinfo{author}{\bibfnamefont{I.~V.} \bibnamefont{Kukushkin}},
	\bibinfo{author}{\bibfnamefont{J.}~\bibnamefont{Falson}},
	\bibinfo{author}{\bibfnamefont{D.}~\bibnamefont{Zhang}},
	\bibinfo{author}{\bibfnamefont{D.}~\bibnamefont{Maryenko}},
	\bibinfo{author}{\bibfnamefont{Y.}~\bibnamefont{Kozuka}},
	\bibinfo{author}{\bibfnamefont{A.}~\bibnamefont{Tsukazaki}},
	\bibinfo{author}{\bibfnamefont{J.~H.} \bibnamefont{Smet}}, \bibnamefont{and}
	\bibinfo{author}{\bibfnamefont{M.}~\bibnamefont{Kawasaki}},
	\bibinfo{journal}{Applied Physics Letters} \textbf{\bibinfo{volume}{106}},
	\bibinfo{pages}{082102} (\bibinfo{year}{2015}),
	\eprint{https://doi.org/10.1063/1.4913313},
	\urlprefix\url{https://doi.org/10.1063/1.4913313}.
	
	\bibitem[{\citenamefont{Zorin}(1995)}]{zorin:1995}
	\bibinfo{author}{\bibfnamefont{A.~B.} \bibnamefont{Zorin}},
	\bibinfo{journal}{Review of Scientific Instruments}
	\textbf{\bibinfo{volume}{66}}, \bibinfo{pages}{4296} (\bibinfo{year}{1995}),
	\eprint{https://doi.org/10.1063/1.1145385},
	\urlprefix\url{https://doi.org/10.1063/1.1145385}.
	
	\bibitem[{\citenamefont{Pines}(2018)}]{pines:2018}
	\bibinfo{author}{\bibfnamefont{D.}~\bibnamefont{Pines}},
	\emph{\bibinfo{title}{Theory of Quantum Liquids: Normal Fermi Liquids}}
	(\bibinfo{publisher}{CRC Press}, \bibinfo{year}{2018}).
	
	\bibitem[{\citenamefont{Tsukazaki et~al.}(2008)\citenamefont{Tsukazaki, Ohtomo,
			Kawasaki, Akasaka, Yuji, Tamura, Nakahara, Tanabe, Kamisawa, Gokmen
			et~al.}}]{tsukazaki:2008}
	\bibinfo{author}{\bibfnamefont{A.}~\bibnamefont{Tsukazaki}},
	\bibinfo{author}{\bibfnamefont{A.}~\bibnamefont{Ohtomo}},
	\bibinfo{author}{\bibfnamefont{M.}~\bibnamefont{Kawasaki}},
	\bibinfo{author}{\bibfnamefont{S.}~\bibnamefont{Akasaka}},
	\bibinfo{author}{\bibfnamefont{H.}~\bibnamefont{Yuji}},
	\bibinfo{author}{\bibfnamefont{K.}~\bibnamefont{Tamura}},
	\bibinfo{author}{\bibfnamefont{K.}~\bibnamefont{Nakahara}},
	\bibinfo{author}{\bibfnamefont{T.}~\bibnamefont{Tanabe}},
	\bibinfo{author}{\bibfnamefont{A.}~\bibnamefont{Kamisawa}},
	\bibinfo{author}{\bibfnamefont{T.}~\bibnamefont{Gokmen}},
	\bibnamefont{et~al.}, \bibinfo{journal}{Phys. Rev. B}
	\textbf{\bibinfo{volume}{78}}, \bibinfo{pages}{233308}
	(\bibinfo{year}{2008}),
	\urlprefix\url{https://link.aps.org/doi/10.1103/PhysRevB.78.233308}.
	
	\bibitem[{\citenamefont{Kozuka et~al.}(2012)\citenamefont{Kozuka, Tsukazaki,
			Maryenko, Falson, Bell, Kim, Hikita, Hwang, and Kawasaki}}]{kozuka:2012a}
	\bibinfo{author}{\bibfnamefont{Y.}~\bibnamefont{Kozuka}},
	\bibinfo{author}{\bibfnamefont{A.}~\bibnamefont{Tsukazaki}},
	\bibinfo{author}{\bibfnamefont{D.}~\bibnamefont{Maryenko}},
	\bibinfo{author}{\bibfnamefont{J.}~\bibnamefont{Falson}},
	\bibinfo{author}{\bibfnamefont{C.}~\bibnamefont{Bell}},
	\bibinfo{author}{\bibfnamefont{M.}~\bibnamefont{Kim}},
	\bibinfo{author}{\bibfnamefont{Y.}~\bibnamefont{Hikita}},
	\bibinfo{author}{\bibfnamefont{H.~Y.} \bibnamefont{Hwang}}, \bibnamefont{and}
	\bibinfo{author}{\bibfnamefont{M.}~\bibnamefont{Kawasaki}},
	\bibinfo{journal}{Phys. Rev. B} \textbf{\bibinfo{volume}{85}},
	\bibinfo{pages}{075302} (\bibinfo{year}{2012}),
	\urlprefix\url{https://link.aps.org/doi/10.1103/PhysRevB.85.075302}.
	
	\bibitem[{\citenamefont{Shashkin et~al.}(2002)\citenamefont{Shashkin,
			Kravchenko, Dolgopolov, and Klapwijk}}]{shashkin:2002}
	\bibinfo{author}{\bibfnamefont{A.~A.} \bibnamefont{Shashkin}},
	\bibinfo{author}{\bibfnamefont{S.~V.} \bibnamefont{Kravchenko}},
	\bibinfo{author}{\bibfnamefont{V.~T.} \bibnamefont{Dolgopolov}},
	\bibnamefont{and} \bibinfo{author}{\bibfnamefont{T.~M.}
		\bibnamefont{Klapwijk}}, \bibinfo{journal}{Phys. Rev. B}
	\textbf{\bibinfo{volume}{66}}, \bibinfo{pages}{073303}
	(\bibinfo{year}{2002}),
	\urlprefix\url{https://link.aps.org/doi/10.1103/PhysRevB.66.073303}.
	
	\bibitem[{\citenamefont{Lilly et~al.}(2003)\citenamefont{Lilly, Reno, Simmons,
			Spielman, Eisenstein, Pfeiffer, West, Hwang, and Das~Sarma}}]{lilly:2003}
	\bibinfo{author}{\bibfnamefont{M.~P.} \bibnamefont{Lilly}},
	\bibinfo{author}{\bibfnamefont{J.~L.} \bibnamefont{Reno}},
	\bibinfo{author}{\bibfnamefont{J.~A.} \bibnamefont{Simmons}},
	\bibinfo{author}{\bibfnamefont{I.~B.} \bibnamefont{Spielman}},
	\bibinfo{author}{\bibfnamefont{J.~P.} \bibnamefont{Eisenstein}},
	\bibinfo{author}{\bibfnamefont{L.~N.} \bibnamefont{Pfeiffer}},
	\bibinfo{author}{\bibfnamefont{K.~W.} \bibnamefont{West}},
	\bibinfo{author}{\bibfnamefont{E.~H.} \bibnamefont{Hwang}}, \bibnamefont{and}
	\bibinfo{author}{\bibfnamefont{S.}~\bibnamefont{Das~Sarma}},
	\bibinfo{journal}{Phys. Rev. Lett.} \textbf{\bibinfo{volume}{90}},
	\bibinfo{pages}{056806} (\bibinfo{year}{2003}),
	\urlprefix\url{https://link.aps.org/doi/10.1103/PhysRevLett.90.056806}.
	
	\bibitem[{\citenamefont{Zabrodskii and
			Zinov'eva}(1984)}]{zabrodskii_low-temperature_1984}
	\bibinfo{author}{\bibfnamefont{A.}~\bibnamefont{Zabrodskii}} \bibnamefont{and}
	\bibinfo{author}{\bibfnamefont{K.}~\bibnamefont{Zinov'eva}},
	\bibinfo{journal}{JETP} \textbf{\bibinfo{volume}{59}}, \bibinfo{pages}{425}
	(\bibinfo{year}{1984}),
	\urlprefix\url{http://www.jetp.ac.ru/cgi-bin/e/index/e/59/2/p425?a=list}.
	
	\bibitem[{\citenamefont{Shashkin et~al.}(2001)\citenamefont{Shashkin,
			Kravchenko, and Klapwijk}}]{shashkin:2001}
	\bibinfo{author}{\bibfnamefont{A.~A.} \bibnamefont{Shashkin}},
	\bibinfo{author}{\bibfnamefont{S.~V.} \bibnamefont{Kravchenko}},
	\bibnamefont{and} \bibinfo{author}{\bibfnamefont{T.~M.}
		\bibnamefont{Klapwijk}}, \bibinfo{journal}{Phys. Rev. Lett.}
	\textbf{\bibinfo{volume}{87}}, \bibinfo{pages}{266402}
	(\bibinfo{year}{2001}),
	\urlprefix\url{https://link.aps.org/doi/10.1103/PhysRevLett.87.266402}.
	
	\bibitem[{\citenamefont{Ando et~al.}(1982)\citenamefont{Ando, Fowler, and
			Stern}}]{ando_electronic_1982}
	\bibinfo{author}{\bibfnamefont{T.}~\bibnamefont{Ando}},
	\bibinfo{author}{\bibfnamefont{A.~B.} \bibnamefont{Fowler}},
	\bibnamefont{and} \bibinfo{author}{\bibfnamefont{F.}~\bibnamefont{Stern}},
	\bibinfo{journal}{Reviews of Modern Physics} \textbf{\bibinfo{volume}{54}},
	\bibinfo{pages}{437} (\bibinfo{year}{1982}), \bibinfo{note}{publisher:
		American Physical Society},
	\urlprefix\url{https://link.aps.org/doi/10.1103/RevModPhys.54.437}.
	
	\bibitem[{\citenamefont{Skinner and Shklovskii}(2013)}]{skinner_theory_2013}
	\bibinfo{author}{\bibfnamefont{B.}~\bibnamefont{Skinner}} \bibnamefont{and}
	\bibinfo{author}{\bibfnamefont{B.~I.} \bibnamefont{Shklovskii}},
	\bibinfo{journal}{Physical Review B} \textbf{\bibinfo{volume}{87}},
	\bibinfo{pages}{075454} (\bibinfo{year}{2013}), \bibinfo{note}{publisher:
		American Physical Society},
	\urlprefix\url{https://link.aps.org/doi/10.1103/PhysRevB.87.075454}.
	
	\bibitem[{\citenamefont{Bello et~al.}(1981)\citenamefont{Bello, Levin,
			Shklovskii, and Efros}}]{Bello_Levin_Shklovskii_Efros_1981}
	\bibinfo{author}{\bibfnamefont{M.~S.} \bibnamefont{Bello}},
	\bibinfo{author}{\bibfnamefont{E.~I.} \bibnamefont{Levin}},
	\bibinfo{author}{\bibfnamefont{B.~I.} \bibnamefont{Shklovskii}},
	\bibnamefont{and} \bibinfo{author}{\bibfnamefont{A.~L.} \bibnamefont{Efros}},
	\bibinfo{journal}{Sov. Phys. JETP} \textbf{\bibinfo{volume}{53}},
	\bibinfo{pages}{822} (\bibinfo{year}{1981}).
	
	\bibitem[{\citenamefont{Eisenstein et~al.}(1992)\citenamefont{Eisenstein,
			Boebinger, Pfeiffer, West, and He}}]{eisenstein:1992}
	\bibinfo{author}{\bibfnamefont{J.~P.} \bibnamefont{Eisenstein}},
	\bibinfo{author}{\bibfnamefont{G.~S.} \bibnamefont{Boebinger}},
	\bibinfo{author}{\bibfnamefont{L.~N.} \bibnamefont{Pfeiffer}},
	\bibinfo{author}{\bibfnamefont{K.~W.} \bibnamefont{West}}, \bibnamefont{and}
	\bibinfo{author}{\bibfnamefont{S.}~\bibnamefont{He}}, \bibinfo{journal}{Phys.
		Rev. Lett.} \textbf{\bibinfo{volume}{68}}, \bibinfo{pages}{1383}
	(\bibinfo{year}{1992}),
	\urlprefix\url{https://link.aps.org/doi/10.1103/PhysRevLett.68.1383}.
	
	\bibitem[{\citenamefont{Li et~al.}(2013)\citenamefont{Li, Zhang, Chong, and
			Hou}}]{li:2013}
	\bibinfo{author}{\bibfnamefont{Q.}~\bibnamefont{Li}},
	\bibinfo{author}{\bibfnamefont{J.}~\bibnamefont{Zhang}},
	\bibinfo{author}{\bibfnamefont{J.}~\bibnamefont{Chong}}, \bibnamefont{and}
	\bibinfo{author}{\bibfnamefont{X.}~\bibnamefont{Hou}},
	\bibinfo{journal}{Applied Physics Express} \textbf{\bibinfo{volume}{6}},
	\bibinfo{pages}{121102} (\bibinfo{year}{2013}),
	\urlprefix\url{https://doi.org/10.7567/apex.6.121102}.
	
	\bibitem[{\citenamefont{Maryenko et~al.}(2018)\citenamefont{Maryenko, McCollam,
			Falson, Kozuka, Bruin, Zeitler, and Kawasaki}}]{maryenko:2018}
	\bibinfo{author}{\bibfnamefont{D.}~\bibnamefont{Maryenko}},
	\bibinfo{author}{\bibfnamefont{A.}~\bibnamefont{McCollam}},
	\bibinfo{author}{\bibfnamefont{J.}~\bibnamefont{Falson}},
	\bibinfo{author}{\bibfnamefont{Y.}~\bibnamefont{Kozuka}},
	\bibinfo{author}{\bibfnamefont{J.}~\bibnamefont{Bruin}},
	\bibinfo{author}{\bibfnamefont{U.}~\bibnamefont{Zeitler}}, \bibnamefont{and}
	\bibinfo{author}{\bibfnamefont{M.}~\bibnamefont{Kawasaki}},
	\bibinfo{journal}{Nature communications} \textbf{\bibinfo{volume}{9}},
	\bibinfo{pages}{4356} (\bibinfo{year}{2018}).
	
	\bibitem[{\citenamefont{Vitkalov et~al.}(2001)\citenamefont{Vitkalov, Zheng,
			Mertes, Sarachik, and Klapwijk}}]{vitkalov:2001}
	\bibinfo{author}{\bibfnamefont{S.~A.} \bibnamefont{Vitkalov}},
	\bibinfo{author}{\bibfnamefont{H.}~\bibnamefont{Zheng}},
	\bibinfo{author}{\bibfnamefont{K.~M.} \bibnamefont{Mertes}},
	\bibinfo{author}{\bibfnamefont{M.~P.} \bibnamefont{Sarachik}},
	\bibnamefont{and} \bibinfo{author}{\bibfnamefont{T.~M.}
		\bibnamefont{Klapwijk}}, \bibinfo{journal}{Phys. Rev. Lett.}
	\textbf{\bibinfo{volume}{87}}, \bibinfo{pages}{086401}
	(\bibinfo{year}{2001}),
	\urlprefix\url{https://link.aps.org/doi/10.1103/PhysRevLett.87.086401}.
	
	\bibitem[{\citenamefont{Kapitulnik et~al.}(2019)\citenamefont{Kapitulnik,
			Kivelson, and Spivak}}]{kapitulnik:2019}
	\bibinfo{author}{\bibfnamefont{A.}~\bibnamefont{Kapitulnik}},
	\bibinfo{author}{\bibfnamefont{S.~A.} \bibnamefont{Kivelson}},
	\bibnamefont{and} \bibinfo{author}{\bibfnamefont{B.}~\bibnamefont{Spivak}},
	\bibinfo{journal}{Rev. Mod. Phys.} \textbf{\bibinfo{volume}{91}},
	\bibinfo{pages}{011002} (\bibinfo{year}{2019}),
	\urlprefix\url{https://link.aps.org/doi/10.1103/RevModPhys.91.011002}.
	
\end{thebibliography}
\end{document}